\documentclass[journal]{IEEEtran}
\usepackage{amssymb}
\usepackage{mathtools}
\usepackage{diagbox}
\usepackage{graphicx,cite,epsfig,amssymb,amsmath}
\usepackage{pifont}
\usepackage{bm}
\usepackage{color}
\usepackage{stmaryrd}        
\usepackage{standalone}
\usepackage{preview}           
\usepackage{xcolor}
\usepackage{bbding}

\usepackage{booktabs}
\usepackage{listings}
\usepackage{multirow}
\usepackage[colorlinks,linkcolor=black,anchorcolor=black,citecolor=black]{hyperref}

\usepackage[makeroom]{cancel} % To cross something
\DeclarePairedDelimiter{\ceil}{\lceil}{\rceil} % for ceiling function
\usepackage{amsthm}
\usepackage{bbm} % for indicator function
\usepackage[caption=false, font=footnotesize]{subfig} % for subfigures
\usepackage{adjustbox}

\usepackage{titlesec}
\newtheorem{theorem}{Theorem}
\newtheorem{lemma}{Lemma}

\usepackage{hyperref}

\UseRawInputEncoding

\begin{document}
% \huge
\title{On the Metric and Computation of PAC Codes}

\author{Mohsen Moradi\textsuperscript{\href{https://orcid.org/0000-0001-7026-0682}{\includegraphics[scale=0.06]{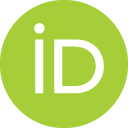}}},~\IEEEmembership{Student Member,~IEEE}% <-this % stops a space
\thanks{The author is with the Department of Electrical-Electronics Engineering, Bilkent University, Ankara TR-06800, Turkey (e-mail: moradi@ee.bilkent.edu.tr).}% <-this % stops a space
}

\maketitle
\begin{abstract}
In this paper, we present an optimal metric function on average, which leads to a significantly low decoding computation while maintaining the superiority of the polarization-adjusted convolutional (PAC) codes' error-correction performance.
With our proposed metric function, the PAC codes' decoding computation is comparable to the conventional convolutional codes (CC) sequential decoding. 
Moreover, simulation results show an improvement in the low-rate PAC codes' error-correction performance when using our proposed metric function. 
We prove that choosing the polarized cutoff rate as the metric function's bias value reduces the probability of the sequential decoder advancing in the wrong path exponentially with respect to the wrong path depth.
We also prove that the upper bound of the PAC codes' computation has a Pareto distribution; our simulation results also verify this.
Furthermore, we present a scaling-bias procedure and a method of choosing threshold spacing for the search-limited sequential decoding that substantially improves the decoder's average computation.  
Our results show that for some codes with a length of 128, the search-limited PAC codes can achieve an error-correction performance close to the error-correction performance of the polar codes under successive cancellation list decoding with a list size of 64 and CRC length of 11 with a considerably lower computation.

%##########################################################################################################

\end{abstract}

%##########################################################################################################

\section{Introduction}

\IEEEPARstart{S}{hort} block length codes with low complexity and probability of error are of concern from a practical perspective. 
Over binary-input additive white Gaussian noise (BI-AWGN) channel, polarization-adjusted convolutional (PAC) codes under sequential decoding \cite{arikan2019sequential}
are shown to have an error-correction performance close to the minimum block error dispersion approximation \cite{polyanskiy2010channel} for a short block length.

It is beneficial to consider the PAC code as a convolutional code (CC) under sequential decoding, which sees a polarized channel.
In this way, the encoding process of PAC codes can be presented by an irregular binary tree, and the sequential decoding of PAC codes can be considered as a walk through a tree.
The decoder's task is to find the correct path in this tree with a guide of a metric functıon. 
Additionally, the channel polarization brings the bit-channel cutoff rates $E_0(1,W_N^{(i)})$ close to the bit-channel capacities $I(W_N^{(i)})$.
We review efficient methods of calculating $E_0(1, W_N^{(i)})$ and $I(W_N^{(i)})$ values in the next section.
We will propose an optimal metric on average, which, by using $E_0(1, W_N^{(i)})$ or $I(W_N^{(i)})$ as the bias values, maintains the superior error-correction performance of the PAC codes while requiring low computation comparable with the computation of the sequential decoding of CCs. 
Compared to the previously used fixed bias values for low code rates \cite{moradi2020PAC}, using either of these bias values improves error-correction performance while keeping low computation of the PAC codes.

By using the metric bias values less than or equal to the bit-channel cutoff rates, we derive an upper bound with Pareto distribution for the computation of the PAC sequential decoder.

Due to the computation variable Pareto distribution, to decode a small fraction of the codewords, the PAC sequential decoder requires high computation. 
Search-limited PAC sequential decoding can be employed \cite{moradi2020PAC} to address this drawback.
We investigate the effect of the threshold spacing $\Delta$ value on the error-correction performance and computation of the search-limited PAC sequential decoder.
Moreover, we propose a scaling method for the bit-channel bias values, which allows employing a trade-off between the error-correction performance and computation of the search-limited PAC sequential decoder.
 
We obtain a good choice of the threshold spacing $\Delta$ value for unlimited search both theoretically and practically.  
Furthermore, we examine the PAC codes' response to code rate to cutoff rate ratio ($R/R_0$).
The simulation results show that, unlike conventional CCs, the PAC codes' computation depends linearly on this ratio. 

Simulation results show that using a fixed bias value in the PAC codes' sequential decoding metric function will result in a satisfactory error performance due to the channel polarization. 
Nevertheless, the computation of the PAC codes' sequential decoding is much higher than the computation of the sequential decoding of CCs \cite{moradi2020PAC}.
We will explain in detail why using a fixed bias value can result in an exponentially high computation.

The decoding computation of the sequential decoder is a random variable.
This paper will represent this random variable by counting the number of nodes that are visited during a single decoding session.

For sufficiently large simulation trials, we use the notion of the average number of visits (ANV), which corresponds to the empirical average of the number of visits. 
We also use the maximum number of visits (MNV) to denote the maximum number of nodes the decoder is allowed to visit during a single decoding session.

Throughout this paper, we will only consider the PAC codes with a codeword of length 128.
All the codes are over the binary Galois field $\mathbb{F}_2 = \{0,1\}.$
We use the boldface notation for the vectors and for a vector $\mathbf{u} = (u_1, u_2, ..., u_N) \in \mathbb{F}_2^N$, $\mathbf{u}^i$ denotes the subvector $(u_1, u_2, ..., u_i)$ and $\mathbf{u}_i^j$ denotes the subvector $(u_i, ..., u_j)$ for $i\leq j$.
For any subset of indices $\mathcal{A} \subset \{1, 2, ..., N\}$, $\mathcal{A}^c$ denotes the complement of $\mathcal{A}$ and $\mathbf{u}_{\mathcal{A}}$ represents the subvector $(u_i : i\in \mathcal{A})$.

The rest of this paper is organized as follows. 
In Section \ref{sec: preliminary}, the calculation of bit-channel mutual information and cutoff rate is explained, and the scheme of the PAC codes is briefly reviewed. 
In Section \ref{sec: metric}, the on average optimality of metric used in the PAC sequential decoder is proved. 
In Section \ref{sec: bias}, the behavior of the partial path (accumulated) metric function is analyzed using different bias values.
In Section \ref{sec: info_frozen simulation}, it is demonstrated that both frozen bits and information bits can be used in the decoding process. 
Moreover, in this Section, it is shown that by using bit-channel capacity or cutoff rate as the metric bias, PAC codes can have an excellent error-correction performance with a constant computation.
In Section \ref{sec: scaling bias} and \ref{sec: threshold spacing}, the techniques for finding the best bias and threshold spacing values both for the search-limited and search-unlimited sequential decoding are proposed.
In Section \ref{sec: Pareto} and Section \ref{sec: upper bound Pareto}, both with simulations and theoretically, it is proved that the computation of the PAC codes has a Pareto distribution upper bound.
Finally, Section \ref{sec: conclusion} concludes this paper with a brief and suggestions for future work.
Appendix \ref{Appendix} provides the proofs of the lemmas and theorems used in this paper.

%##########################################################################################################

\section{Preliminaries}\label{sec: preliminary}
This section will review the efficient methods of calculating bit-channel mutual information and the bit-channel cutoff rate, which we will frequently use in this paper.

\subsection{Bit-channel mutual information}
Consider a BI-AWGN channel with binary phase-shift keying (BPSK) modulation. 
If the channel input is a uniform random variable $X \in \{-1, +1 \}$ and the channel output is a random variable $Y \in \mathbb{R}$, the mutual information between the input and the output is defined as 

\begin{equation} \label{initialization}
 \begin{split}
     I(W) & = I(X;Y)    \\
     & := \sum_{x = \pm 1}\int_{- \infty}^{+ \infty} \frac{1}{2} P_{Y|X}(y|x) \log_2\frac{P_{Y|X}(y|x)}{P_{Y}(y)} dy  \\
     & = \frac{1}{\sqrt{8\pi\sigma^2}}\int_{- \infty}^{+ \infty} a\left(1-\log_2(\frac{b}{a}+1)\right) \ + \\
     & \ \ \ \ \ \ \ \ \ \ \ \ \ \ \ \ \ \ \ \ \ \ b\left(1-\log_2(\frac{a}{b}+1)\right)dy,
 \end{split}   
\end{equation} 
where
\begin{equation}
    a := e^{\frac{-(y-1)^2}{2\sigma^2}} \text{     and     } b := e^{\frac{-(y+1)^2}{2\sigma^2}},
\end{equation}
and $\sigma^2$ is the noise variance of the BI-AWGN channel.
Since $y = x + n$ has a Gaussian distribution, where $n$ is a Gaussian noise with zero mean and variance $\sigma_{n}^2$, according to \cite{brannstrom2005convergence} 
\begin{equation} 
    I(X;Y) = J(\frac{2}{\sigma_{n}^2}),
\end{equation}
where
\begin{equation}
    J(t) = 1 - \frac{1}{\sqrt{2\pi t^2}} \int_{- \infty}^{+ \infty} e^{\frac{-(u-\frac{t^2}{2})^2}{2t^2}}\log_2(1+e^{-u})du.
\end{equation}
The monotonically increasing function $I(X;Y) = J(t)$ has a unique inverse function $t = J^{-1}(I(X;Y)).$ 
An approximation for the function $J(t)$ and its inverse is given in \cite{brannstrom2004convergence} as
\begin{equation}
    \begin{split}
        & J(t) = \left[1 - 2^{-0.3073 t^{2\times0.8935}} \right]^{1.1064},\\
        & J^{-1}(I(X;Y)) = \\
        & \ \ \ \ \ \left[-\frac{1}{0.3073}\log_2(1-I(X;Y)^{\frac{1}{1.1064}}) \right]^\frac{1}{2\times 0.8935}.
    \end{split}
\end{equation}

\begin{figure}[htbp]
\centering
	\includegraphics [width = \columnwidth]{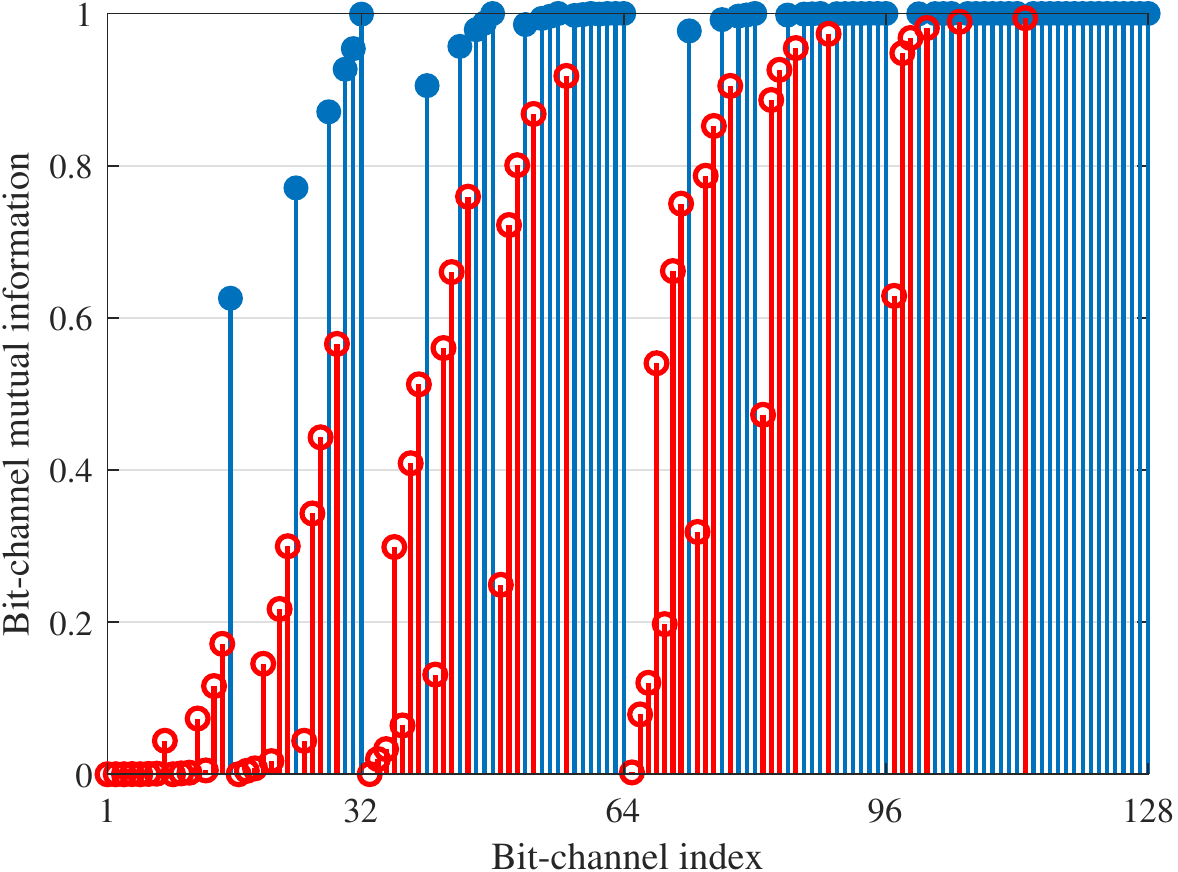}
	\caption{Bit-channel capacity for information bits (solid circles) and frozen bits (hollow circles) at 2.5~dB SNR value.} 
	\label{fig: bit channel IW}
\end{figure}

In every step of polarization process, independent copies of channel $W$ is transformed into polarized binary input channels $W^+$ and $W^-$. 
If we represent this operation by a tree with its root being initialized by (\ref{initialization}), at each node the mutual information is polarized using the left-branch operation (parity operation) $f_c$ and the right-branch operation (node operation) $f_v$, which are defined as
\begin{equation}
    \begin{split}
        & f_c(t) = 1 - J\left[\sqrt{2}J^{-1}(1-t)\right], \\
        & f_v(t) = J\left[\sqrt{2}J^{-1}(t)\right],
    \end{split}
\end{equation}
where the leaf nodes are the bit-channel mutual information $I(W_{N}^{(i)})$ for $i \in \{1,..., N\}.$ 
Fig. \ref{fig: bit channel IW} plots the bit-channel mutual information for length of 128 and rate of $1/2$ at $\text{SNR} = 2.5$~dB. 
Information bit locations are chosen according to Reed-Muller rate profile (rows with higher weights) as explained in \cite{arikan2019sequential}.

Another important parameter of the channel $W$ is the Bhattacharyya parameter which is defined as
\begin{equation}
    Z(W) := \sum_{y \in \mathcal{Y}} \sqrt{W(y|0)W(y|1)} .
\end{equation}
For a BI-AWGN channel, $Z(W)$ is 
\begin{equation}
    Z(W) = e^{\frac{-1}{2\sigma^2}},
\end{equation}
where $\sigma^2$ is the noise variance of the BI-AWGN channel.  
The log-likelihood ratio of the output of the channel $W$ has a Gaussian distribution with mean $m_0^{(1)} = 2/\sigma^2$ and variance $2m_0^{(1)}$. 
With $m_0^{(1)}$ at the root of tree, the bit-channel means $m_N^{(i)}$ at the leaf level of the tree can be calculated using the check-operations $f_c$ and bit-operation $f_v$ as \cite{li2013practical}

\begin{equation}
    \begin{split}
        & f_c(t) = \phi^{-1}\Big( 1- (1-\phi(t))^2 \big), \\
        & f_v(t) = 2\phi(t),
    \end{split}
\end{equation}
where
\begin{equation}
    \phi(t) =
\begin{cases}
      1-\frac{1}{\sqrt{4\pi t}} \int_\mathbb{R} \tanh(\frac{z}{2}) e^{-\frac{(z-t)^2}{4t}}dz , & t > 0, \\
      1, & t < 0,
\end{cases}
\end{equation}
 and $\phi^{-1}(t)$ can be calculated numerically.

Finally, the bit-channel Bhattacharya parameters can be calculated as
\begin{equation}
    Z(W_N^{(i)}) = e^{\frac{-1}{2(\sigma_N^{(i)})^2}},
\end{equation}
where $m_N^{(i)} = 2/(\sigma_N^{(i)})^2 $.

\begin{figure}[htpb]
\centering
	\includegraphics [width = \columnwidth]{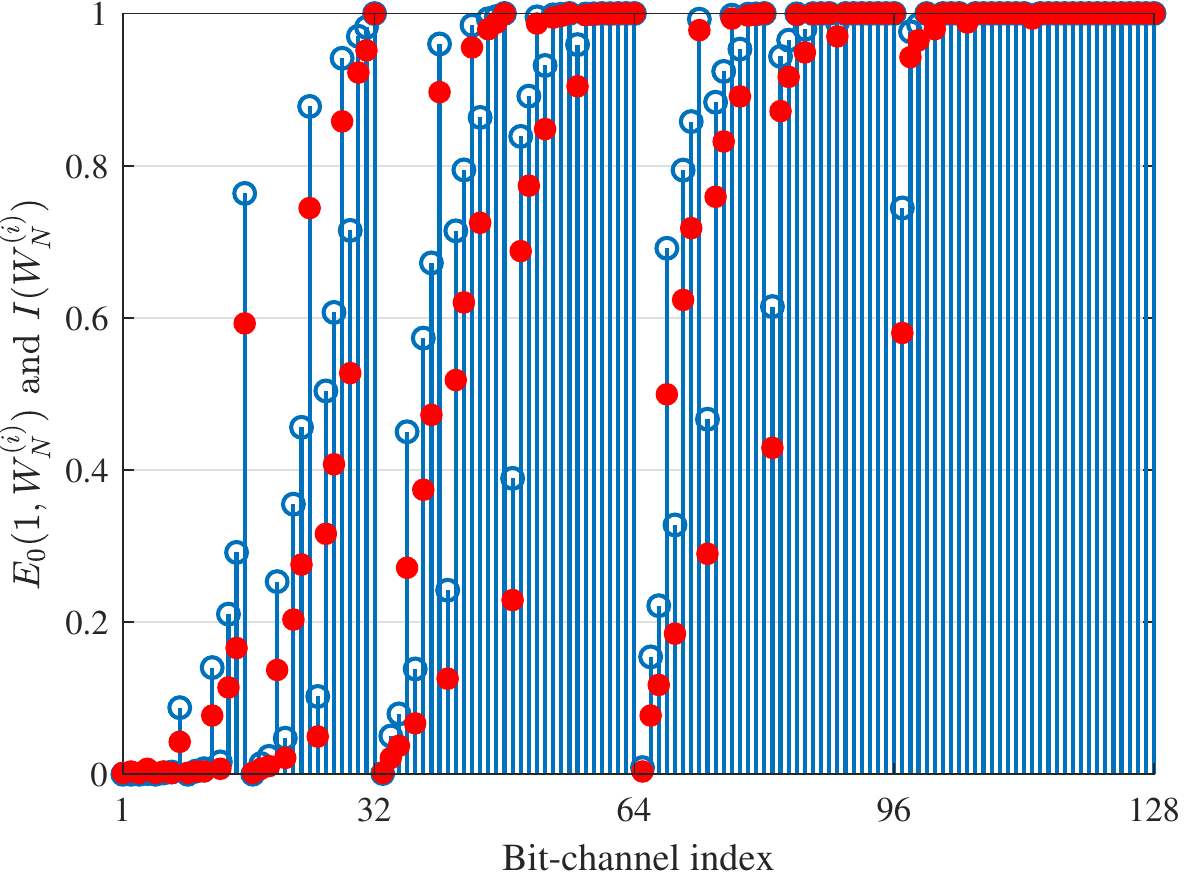}
	\caption{(solid circles) bit-channel cutoff rate and (hollow circles) bit-channel mutual information. } 
	\label{fig: error exponent}
\end{figure}

For a given B-DMC $W$ and any $\rho \geq 0$, the error exponent of the channel $W$ with $q(x)$ distribution on the input is defined as \cite{gallager1968information}
\begin{equation}\label{eq: error exponent}
    E_0(\rho, W) = - \log_2 \sum_{y \in \mathcal{Y}} \left[ \sum_{x \in \mathcal{X}} q(x) W(y|x)^{\frac{1}{1+\rho}}\right] ^{1+\rho} .
\end{equation}
By substituting $\rho = 1$ in \eqref{eq: error exponent}, we can obtain the cutoff rate as
\begin{equation}
    R_0(W,q) := E_0(1, W).
\end{equation}
If the input distribution is uniform, the error exponent becomes
\begin{equation}\label{E0vZW}
   E_0(1,W) = \log_2 \frac{2}{1+Z(W)}, 
\end{equation}
which is a lower bound for $I(W).$ 
According to \eqref{E0vZW}, polarization of $Z(W)$ results in the polarization of $E_0(1,W)$.
We will refer to $E_0(1,W_N^{(i)})$ as bit-channel cutoff rate.
Fig. \ref{fig: error exponent} compares the bit-channel cutoff rate with bit-channel mutual information. 

In \cite{alsan2014polarization}, it is also proven that the error exponent is polarized for any arbitrary $\rho \geq 0$ as
\begin{equation}
    E_0(\rho,W^-) + E_0(\rho,W^+) \geq 2E_0(\rho,W).
\end{equation}

%##########################################################################################################
\begin{figure}[t] 
\centering
	\includegraphics [width = \columnwidth]{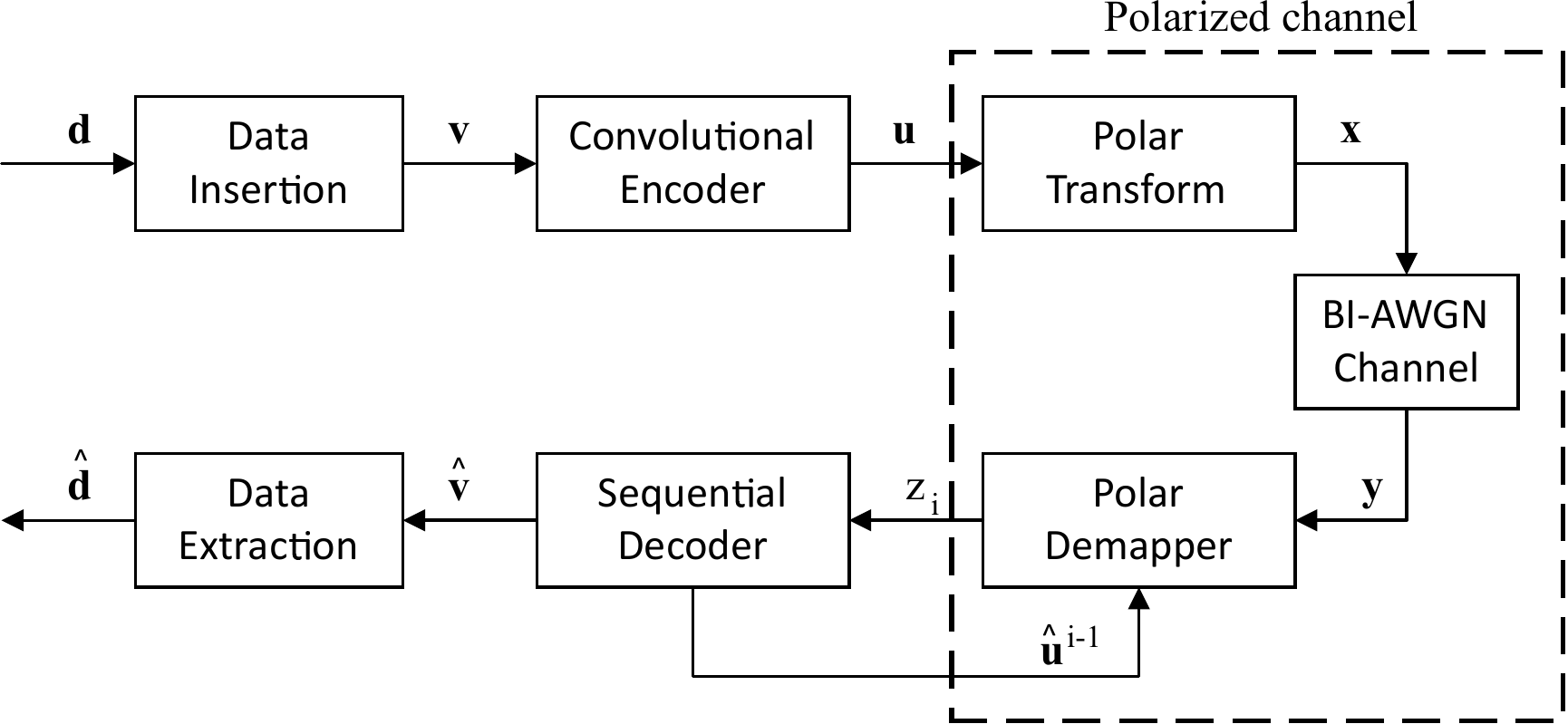}
	\caption{Flowchart of PAC coding scheme.} 
	\label{fig: flowchart}
\end{figure}

\subsection{PAC Coding Scheme} \label{sec: scheme}
Fig. \ref{fig: flowchart} shows a flow chart of the PAC coding scheme.
$\mathbf{d}=(d_1,\ldots,d_K)$ is the data generated uniformly at random over all possible source words of length $K$ in a binary field.
The rate profile module maps these $K$ bits into a data carrier vector $\mathbf{v}$ in accordance with the data set $\mathcal{A}$, thus inducing a code rate of $R=K/N$.
In this paper, the RM score function is employed to determine $\mathcal{A}$ set.  
After $\mathbf{v}$ is obtained by $\mathbf{v}_\mathcal{A} = \mathbf{d}$ and $\mathbf{v}_{\mathcal{A}^c} = 0$, it is sent to the convolutional encoder and encoded as $\mathbf{u} = \mathbf{v}\mathbf{T}$, where $\mathbf{T}$ is an upper-triangular Toeplitz matrix constructed with a connection polynomial $\mathbf{c}$.
In all of our simulations, we use the connection polynomial $\mathbf{c} = 3211$ (in octal form) which is introduced in \cite{moradi2020PAC}.
Then $\mathbf{u}$ is transformed to $\mathbf{x}$ with standard polar transformation $\mathbf{F}^{\otimes n}$, where $\mathbf{F}^{\otimes n}$ is the kronecker power of the kernel matrix $\mathbf{F} = \begin{bsmallmatrix} 1 & 0\\ 1 & 1 \end{bsmallmatrix}$ with $n = \log_2 N$.
After the polar transformation, $\mathbf{x}$ is sent through the BI-AWGN channel.
Polar demapper receives the channel output $\mathbf{y}$ and the previously decoded bits and calculates the log-likelihood ratio (LLR) value of the current bit $z_i$.
Finally, the sequential decoder outputs an estimate of the carrier word $\hat{\mathbf{v}}$, from which the $K$-bits data can be extracted according to $\mathcal{A}$.

The channel that the sequential decoder of PAC codes sees is a polarized channel with memory.
For this reason, the metric used by the sequential decoder must be compatible with the channel.
In the following sections, we will investigate the metric function. 
To implement the sequential decoder, we use the Fano algorithm  \cite{fano1963heuristic}, which is modified to be compatible with the PAC codes \cite{mozammel2020hardware}. 
Two parameters of the Fano algorithms are the bias and the threshold spacing.
In the rest of this paper, we will investigate the effects of bias and threshold spacing on error-correction performance and computation of the PAC codes under search-limited and search-unlimited decoder both theoretically and practically.    

%##########################################################################################################

\section{Metric}\label{sec: metric}
The most commonly used metric for sequential decoding in a B-DMC is Fano metric \cite{fano1963heuristic} and is defined as
\begin{equation}
    \gamma(x_i;y_i) = \log_2 \frac{P(y_i | x_i)}{P(y_i)} - b,
\end{equation}
where $P(y_i | x_i)$ is the channel transition probability, $P(y_i)$ is the channel output, and $b$ is a constant bias.
Massey \cite{massey1972variable} shows that for a B-DMC, the Fano metric is the optimum metric for comparing paths of different lengths.
But clearly, the polarized channel in PAC codes is a channel with memory since, for a polarized channel, demapping of any bit requires the values of all the previously decoded bits. 
The Fano metric needs to be modified to be suitable for decoding an irregular tree code sent over a channel with memory. 
By adopting the notation used in \cite{arikan2020PAC},\cite{moradi2020PAC} and recalling that the rate one convolution operation and polar transform are one-to-one transforms, the partial path metric for the first $i$ branches is given by
\begin{equation} \label{partialmetric}
    \Gamma(\mathbf{u}^i;\mathbf{y}) = \log_2 \left( \frac{P(\mathbf{y} | \mathbf{u}^{i})}{P(\mathbf{y})}\right) - B_i,
\end{equation}
where $\mathbf{y} = (y_1, y_2, ..., y_N)$ is the channel output, $\mathbf{u}^i = (u_1, u_2, ..., u_i) $ is the path vector from root of the tree to the node $i$, and $B_i = \sum_{j=1}^ib_j$ is the partial path bias value up to the $i$th bit, and $b_j$ is a design parameter.

Every time a new branch in the tree is being examined, it is more convenient to calculate the branch metric rather than calculating the partial path metric of (\ref{partialmetric}). 
For decoding $u_j$, the decoder knows the channel output $\mathbf{y}$ and the previous bits $u_1$ up to $u_{j-1}$. 
By defining $\gamma(u_j ;\mathbf{y},\mathbf{u}^{j-1})$ as the $j$th branch metric, equation (\ref{partialmetric}) can be written as
\begin{equation}
    \Gamma(\mathbf{u}^{i};\mathbf{y}) = \sum_{j=1}^{i} \gamma(u_j ;\mathbf{y},\mathbf{u}^{j-1}).
\end{equation}

By defining $b_j := B_j - B_{j-1} $ we have
\begin{equation}\label{bitmetric}
\begin{split}
     &\gamma(u_j ;\mathbf{y},\mathbf{u}^{j-1}) = \Gamma(\mathbf{u}^{i};\mathbf{y}) - \Gamma(\mathbf{u}^{j-1};\mathbf{y})  \\
    & =  \left[ \log_2 \left(\frac{P(\mathbf{y} | \mathbf{u}^{i})}{P(\mathbf{y})}\right) - B_j  \right]\\
     & \ \ \ \ - \left[ \log_2 \left(\frac{P(\mathbf{y} | \mathbf{u}^{j-1})}{P(\mathbf{y})}\right) - B_{j-1} \right]
     \\
    & = \log_2\left(\frac{P(\mathbf{y} | \mathbf{u}^{j})}{P(\mathbf{y} | \mathbf{u}^{j-1})}\right) - (B_j - B_{j-1})
     \\
    & = \log_2 \left(\frac{P(\mathbf{y} | \mathbf{u}^{j})}{P(\mathbf{y} | \mathbf{u}^{j-1})}\right) - b_j
     = \log_2 \left(\frac{P(\mathbf{y},\mathbf{u}^{j-1}|u_j)}{P(\mathbf{y},\mathbf{u}^{j-1})}\right) - b_j.
\end{split}
\end{equation}
For a binary input channel with uniform input distribution, $u_j$ can be either 0 or 1. For $u_j = 0$, (\ref{bitmetric}) becomes
\begin{equation}\label{bitzero}
\begin{split}
    & \gamma(u_{j} = 0 ;\mathbf{y},\mathbf{u}^{j-1}) = \log_2 \left(\frac{P(\mathbf{y},\mathbf{u}^{j-1}|u_{j} = 0 )}{P(\mathbf{y},\mathbf{u}^{j-1})}\right) - b_{j}
    \\
    & = \log_2 \left(\frac{P(\mathbf{y},\mathbf{u}^{j-1}|u_{j} = 0 )}{\frac{1}{2} \left[P(\mathbf{y},\mathbf{u}^{j-1}|u_{j} = 0) +P(\mathbf{y},\mathbf{u}^{j-1}|u_{j} = 1)  \right] }\right)\\
    & \ \ \ - b_{j}.
\end{split}
\end{equation}
Dividing the numerator and denominator of (\ref{bitzero}) by $P(\mathbf{y},\mathbf{u}^{j-1}|u_j = 0 )$ and defining 
\begin{equation}
    z_j := \dfrac{P(\mathbf{y},\mathbf{u}^{j-1}|u_j = 0 )}{P(\mathbf{y},\mathbf{u}^{j-1}|u_j = 1 )}
\end{equation}
as the likelihood ratio of bit $u_j$, we can rewrite (\ref{bitzero}) as
\begin{equation}
    \gamma(u_j = 0 ;\mathbf{y},\mathbf{u}^{j-1}) = 1 - \log_2 \left(1 + \dfrac{1}{z_j}\right) - b_j.
\end{equation}
Similarly, we can obtain the $j$th branch metric for $u_j = 1$ as
\begin{equation}
    \gamma(u_j = 1 ;\mathbf{y},\mathbf{u}^{j-1}) = 1 - \log_2 \left(1 + z_j\right) - b_j.
\end{equation}
In conclusion, the $j$th branch metric for an irregular tree code transmitted over a polarized channel is calculated using
\begin{equation}
\label{eqn:bit metric}
\begin{split}
    & \gamma(u_j ;\mathbf{y},\mathbf{u}^{j-1})= \\
    & \ \ \ \
\begin{cases}
      1 - \log_2 \left(1 + \dfrac{1}{z_j}\right) - b_{j},& \text{if } u_{j} = 0; \\
      1 - \log_2(1 + z_j) - b_{j}, & \text{if } u_{j} = 1.
\end{cases}
\end{split}
\end{equation}

% To understand what values of $b_j$ in average make progress in the correct path and make a decrement for the metric for an incorrect path, consider the branch metric by (\ref{bitmetric}),

The following expectations mentioned in \cite{arikan2020PAC} are useful in the next section. 
The expectation of \eqref{bitmetric} over the ensemble of input bit $u_j$ and the bit-channel transition probability $P(\mathbf{y},\mathbf{u}^{j-1} | u_j)$ is
\begin{equation}
    \label{eqn:exp_pos}
    \begin{split}
        & \mathbb{E}_{u_j,\left(\mathbf{y},\mathbf{u}^{j-1} \right)} [\gamma(u_j ;\mathbf{y},\mathbf{u}^{j-1})]
        \\
        & = \sum_{u_j}q(u_j) \sum_{\mathbf{y},\mathbf{u}^{j-1}} P(\mathbf{y},\mathbf{u}^{j-1}|u_j) \gamma(u_j ;\mathbf{y},\mathbf{u}^{j-1})
        \\
       & = \sum_{u_j}\sum_{\mathbf{y},\mathbf{u}^{j-1}}q(u_j)P(\mathbf{y},\mathbf{u}^{j-1}|u_j)
       \\
       &\ \ \ \ \left[ \log_2 \left(\frac{P(\mathbf{y},\mathbf{u}^{j-1}|u_j)}{P(\mathbf{y},\mathbf{u}^{j-1})}\right) - b_j \right]
        \\
        & ذ= I(W_N^{(j)}) - b_j,
    \end{split}
\end{equation}
where $I(W_N^{(j)})$ is the symmetric capacity of the bit-channel. If the expectation in \eqref{eqn:exp_pos} is positive, the average branch metric increment is always positive. Choosing the bit-channel bias less than symmetric capacity of the bit-channels would guarantee that the expectation in \eqref{eqn:exp_pos} is positive. 

Now, assume that $\Tilde{u}_j$ is an incorrect branch with a metric of $ \gamma (\Tilde{u}_j ; \mathbf{y}, \mathbf{u}^{j-1}) $ on the decoding tree. By taking the average on all incorrect choices we can obtain
\begin{equation}
    \label{eqn:exp_neg}
    \begin{split}
        & \mathbb{E}_{u_{j}, \Tilde{u}_j,\left(\mathbf{y},\mathbf{u}^{j-1} \right)} [\gamma(\Tilde{u}_j ;\mathbf{y},\mathbf{u}^{j-1})]
        \\
        & = \sum_{\Tilde{u}_j} q(\Tilde{u}_j) \sum_{u_{j}}q(u_{j}) \sum_{\mathbf{y},\mathbf{u}^{j-1}} P(\mathbf{y},\mathbf{u}^{j-1}|u_{j})
        \gamma(\Tilde{u}_j ;\mathbf{y},\mathbf{u}^{j-1})
        \\
        & =\sum_{\Tilde{u}_j} q(\Tilde{u}_j) \sum_{\mathbf{y},\mathbf{u}^{j-1}} P(\mathbf{y},\mathbf{u}^{j-1}) \gamma(\Tilde{u}_j ;\mathbf{y},\mathbf{u}^{j-1})
        \\
        & = \sum_{\Tilde{u}_j}\sum_{\mathbf{y},\mathbf{u}^{j-1}}q(\Tilde{u}_j)P(\mathbf{y},\mathbf{u}^{j-1})
        \left[ \log_2 \left(\frac{P(\mathbf{y},\mathbf{u}^{j-1}|\Tilde{u}_j)}{P(\mathbf{y},\mathbf{u}^{j-1})}\right) - b_{j} \right]
        \\
        & \leq \sum_{\Tilde{u}_j}\sum_{\mathbf{y},\mathbf{u}^{j-1}}q(\Tilde{u}_j)P(\mathbf{y},\mathbf{u}^{j-1}) \left[ \frac{P(\mathbf{y},\mathbf{u}^{j-1}|\Tilde{u}_j)}{P(\mathbf{y},\mathbf{u}^{j-1})} - 1  \right]  - b_{j}
        \\
        & =  - b_{j}.
    \end{split}
\end{equation}

Note that for deriving \eqref{eqn:exp_neg}, we used the property $\log_2(x) \leq x-1$. 
Since the information bit bias value is positive, the above expectation shows that the branch metric will decrease by a constant value for an incorrect branch on average. 

\subsection{Optimality of the Proposed Metric}
In the $i$th step of decoding, given the channel output $\mathbf{y}$, the optimal metric is obtained when $p(\mathbf{u}^{i}|\mathbf{y})$ is maximized. This probability is calculated by 
\begin{equation}
    p(\mathbf{u}^{i}|\mathbf{y}) = \frac{p(\mathbf{y}|\mathbf{u}^{i})}{p(\mathbf{y})}p(\mathbf{u}^{i}).
\end{equation}

By taking the $\log_2$ function of both sides, the partial path metric from root to node $i$ is defined as
\begin{equation}
    \Gamma (\mathbf{u}^{i}; \mathbf{y}) := \log_2 p(\mathbf{u}^{i}|\mathbf{y}) = \log_2 \frac{p(\mathbf{y}|\mathbf{u}^{i})}{p(\mathbf{y})} + \log_2 p(\mathbf{u}^{i}).
\end{equation}
Note that, since $\log_2$ is a monotonically increasing function, it will preserve maximality.
The $j$th branch metric will become
\begin{equation}
    \begin{split}
        & \gamma(u_j ;\mathbf{y},\mathbf{u}^{j-1}) := \Gamma(\mathbf{u}^{i};\mathbf{y}) - \Gamma(\mathbf{u}^{j-1};\mathbf{y})  \\
        & =   \log_2 \frac{P(\mathbf{y} | \mathbf{u}^{i})}{P(\mathbf{y} | \mathbf{u}^{j-1})} + \log_2 \frac{p(\mathbf{u}^{i})}{p(\mathbf{u}^{j-1})} \\
        & =   \log_2 \frac{P(\mathbf{y},\mathbf{u}^{j-1} | u_{j})}{P(\mathbf{y},\mathbf{u}^{j-1})} + \log_2 \frac{p(\mathbf{u}^{i})}{p(\mathbf{u}^{j-1})} \\
        & =   \log_2 \frac{P(\mathbf{y},\mathbf{u}^{j-1} | u_{j})}{P(\mathbf{y},\mathbf{u}^{j-1})} + \log_2 p(u_j),
    \end{split}
\end{equation}
where the last equality is obtained using the fact that convolution is a one-to-one and deterministic. Defining the bias value as
\begin{equation}
    b_j = \log_2 \frac{1}{p(u_j)},
\end{equation}
$b_j$ would become the information provided about the event $u_j$, which is $I(W_N^{(i)})$ on average.  
Note that due to the polarization effect, the bit-channel capacities are polarized.
For the PAC codes constructed using RM rate profile, there are bit-channel capacities with large values for both frozen and information sets. This paper aims to benefit from these capacities by designing a proper rule for $b_j$.

%%%%%%%%%%%%%%%%%%%%%%%%%%%%%%%%%%%%%%%%%%%%%%%%%%%%%%%%%%%%%%%%%%%%%%%%%%%%%%%%%%%%%%%%%%%%%%%%%%%%%%%%%%%%%%%%%%%%%%%%%%%%%%%%%%%%%%

\section{Design rules for bias}\label{sec: bias}
This section will study the effect of the different possible bias values feasible to use in the sequential decoding of the PAC codes.  
Various choices of bias values can result in different error-correction performance and computation.

We will use the notations $b_i^{(F)}$ and $b_i^{(I)}$ to represent the bias values for the frozen and information bits, respectively.
We will use $b_i$ to indicate the bit-channel bias values whenever we do not specify whether a bit is frozen or information.

For a sequential decoder, to have better error-correction performance and lower computation, the codewords must have a distance that increases as early as possible.
An optimum choice of the bit-channel bias values may result in a fair distance.
As discussed in the last section, by taking an average for the correct and wrong paths, we have:
\begin{equation}\label{eq: av. ineq. bias}
\begin{split}
    &\textit{Correct path:}~~~~~~~~~~ b_i \leq I(W_N^{(i)}),\\
    &\textit{Wrong path:  }~~~~~~~ -b_i \leq 0.\\
\end{split}
\end{equation}
In the following subsections, we will discuss three different design rules for the bit-channel bias values. 
All three rules keep the PAC codes' superior error performance, but their computation is entirely different.

%%%%%%%%%%%%%%%%%%%%%%%%%%%%%%%%%%%%%%%%%%%%%%%%%%%%%%%%%%%%%%%%%%%%%%%%%%%%%%%%%%%%%%%%%%%%%%%%%%%%%%%%%%%%%%%%%%%%%%%%%%%%%%%%%%%%%%

\subsection{Design rule 1}
A metric with the bias values $b_i^{(F)} = 0$ and fixed $b_i^{(I)}$ s.t. $b_i^{(I)} > I(W_N^{(i)})$ is used in \cite{arikan2020PAC} and \cite{moradi2020PAC}. 
More specifically, \cite{moradi2020PAC} uses $b_i^{(I)}$ of 1.4, 1.35, and 1.14 for K = 29, 64, 99, respectively. 
Smaller $b_i^{(I)}$ values will sacrifice the PAC code's error-correction performance.
On the other hand, larger $b_i^{(I)}$ values increase the computation. 
Note that for $i \in \mathcal{A}^c$, $b_i^{(F)} = 0 < I(W_N^{(i)})$ and hence both inequalities of (\ref{eq: av. ineq. bias}) are satisfied.
As a result, the decoder can decode the frozen bits faster, and the partial path metric becomes a positive value for the frozen bits. 
On the other hand, by assigning a large value to the information-bit bias values (say $b_i^{(I)} = 1.35$ for $K = 64$), since $b_i^{(I)} \nleq I(W_N^{(i)}) $, the decoder will have some difficulty to find the correct path, and the partial path metric for the information bits becomes negative on average. 

\begin{figure}[t]
\centering
	\includegraphics [width = \columnwidth]{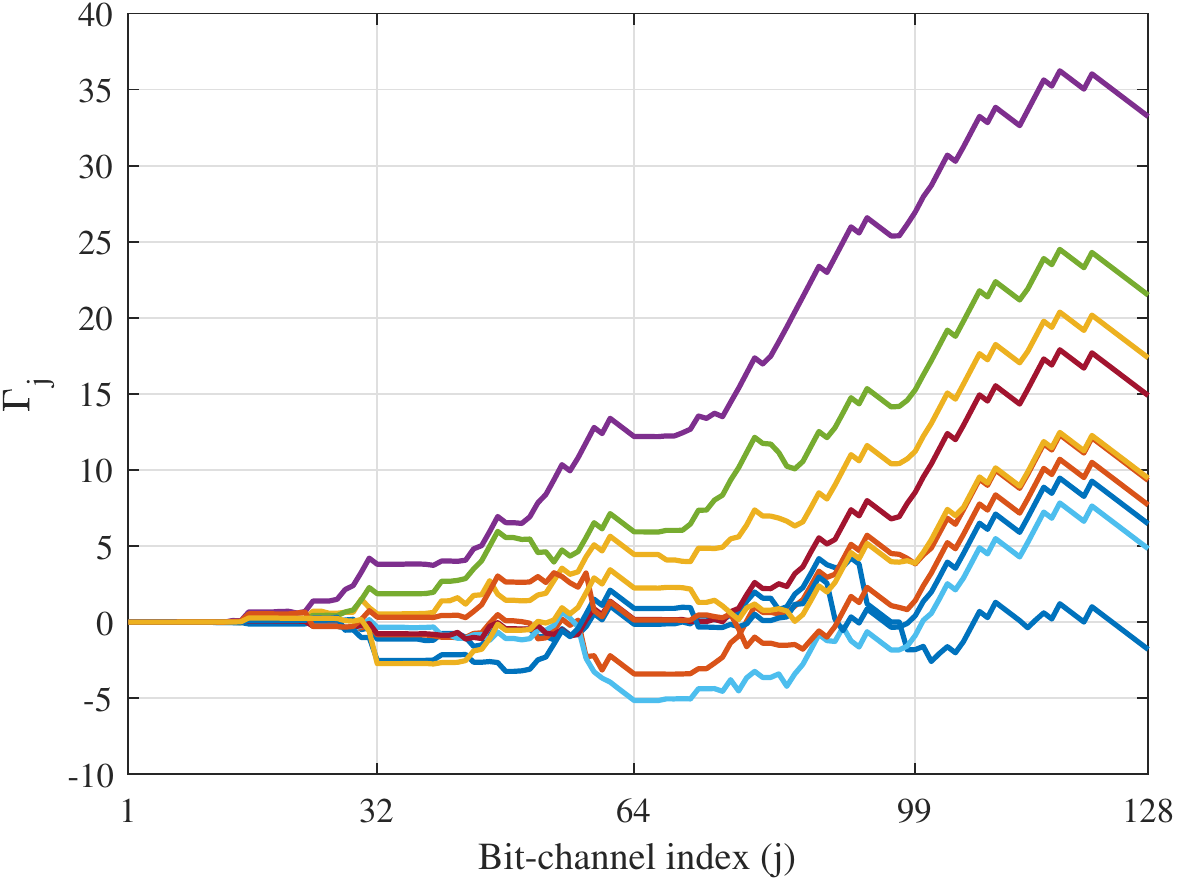}
	\caption{10 sample partial path metrics v. bit indices for K = 29 with $b_i^{(I)} = 1.4$ and $b_i^{(F)} = 0$ at -1.5~dB SNR value.} 
	\label{fig: K29 fixedBI}
\end{figure}

\begin{figure}[h]
\centering
	\includegraphics [width = \columnwidth]{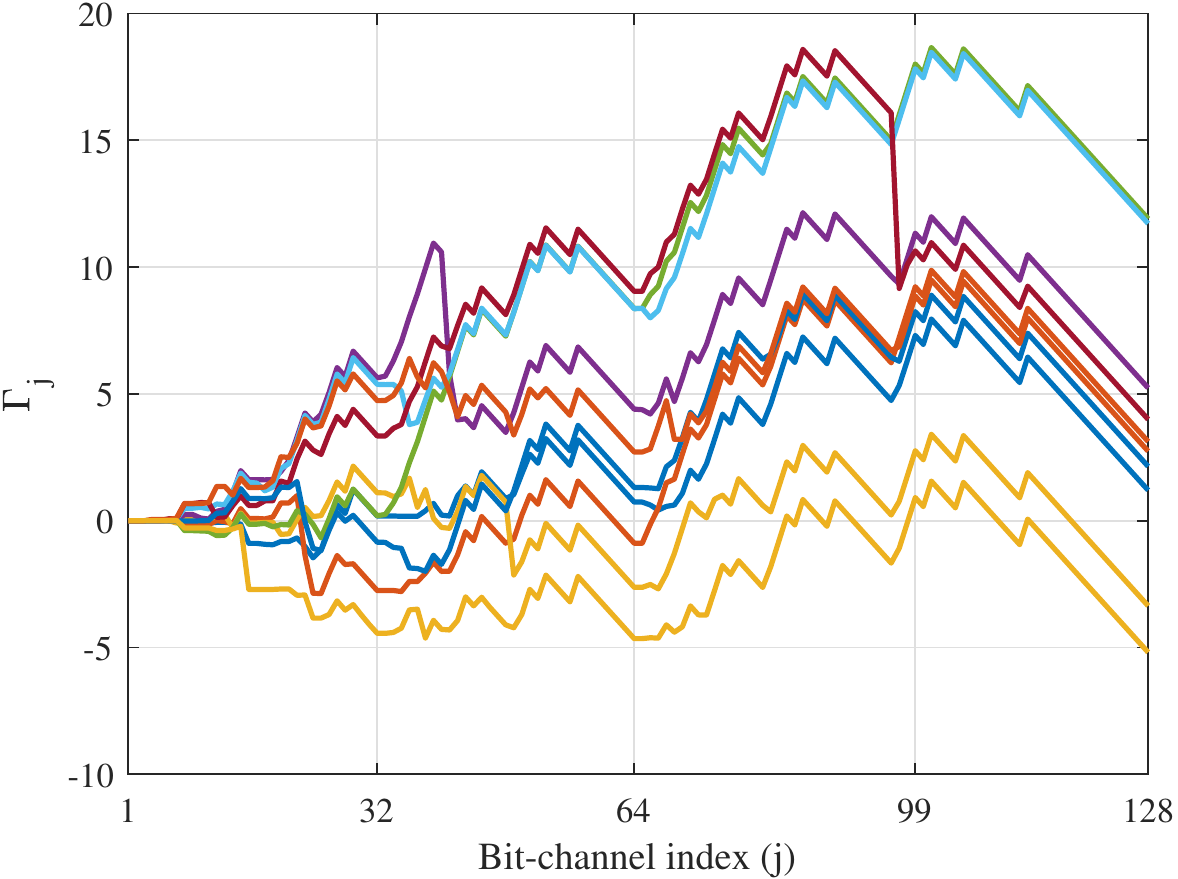}
	\caption{10 sample partial path metrics v. bit indices for K = 64 with $b_i^{(I)} = 1.35$ and $b_i^{(F)} = 0$ at 2.5~dB SNR value.} 
	\label{fig: K64 fixedBI}
\end{figure}

\begin{figure}[h]
\centering
	\includegraphics [width = \columnwidth]{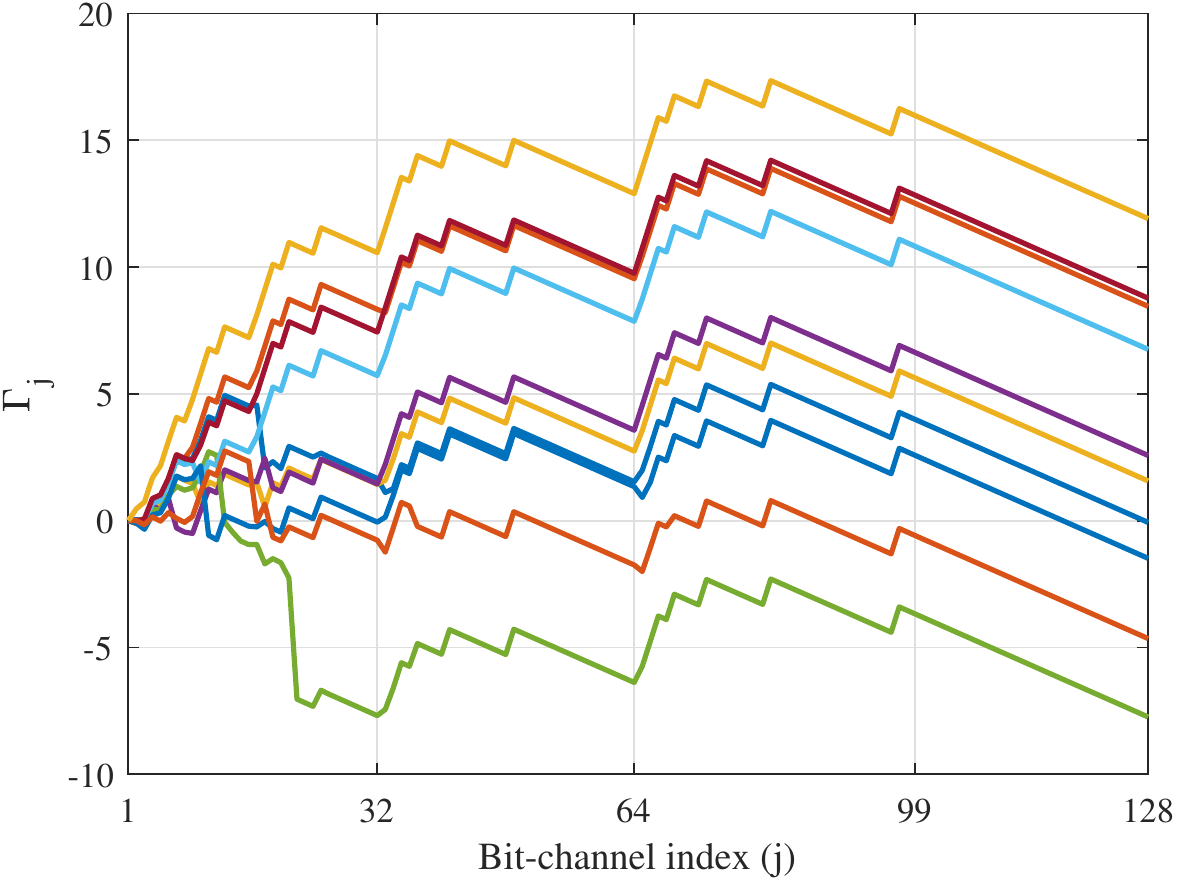}
	\caption{10 sample partial path metrics v. bit indices for K = 99 with $b_i^{(I)} = 1.14$ and $b_i^{(F)} = 0$ at 5.5~dB SNR value.} 
	\label{fig: K99 fixedBI}
\end{figure}

One drawback of choosing $b_i^{(I)} > I(W_N^{(i)})$ is that, if the code has long consecutive information bits, the information bits' partial path metric will significantly fall. 
In \cite{moradi2020PAC}, for $K = 29, 64, 99$ fixed bias values for the information bits are chosen as $b_i^{(I)} = 1.4, 1.35, 1.14$, respectively. 
The reason to choose smaller $b_i^{(I)}$ values for longer messages is to reduce the effect of negative partial path metric; otherwise, the decoding would require an exponentially larger computation. 
Fig. \ref{fig: K29 fixedBI}, \ref{fig: K64 fixedBI}, and \ref{fig: K99 fixedBI} show the partial path metric for these three aforementioned code rates with the corresponding bias values. 
These figures demonstrate that the partial path metric will increase for the frozen bits and will decrease for the information bits on average by choosing a fixed $b_i^{(I)} > I(W_N^{(i)})$ and $b_i^{(F)} = 0 < I(W_N^{(i)})$.

The advantage of choosing $b_i^{(F)} = 0$ is that in this case, both inequalities of (\ref{eq: av. ineq. bias}) are satisfied for the frozen bits, and consequently, the decoder can decode the frozen bits with lower computation.
As long as $I(W_N^{(i)}) > b_i^{(F)}$, the partial path metric is positive on average.

%%%%%%%%%%%%%%%%%%%%%%%%%%%%%%%%%%%%%%%%%%%%%%%%%%%%%%%%%%%%%%%%%%%%%%%%%%%%%%%%%%%%%%%%%%%%%%%%%%%%%%%%%%%%%%%%%%%%%%%%%%%%%%%%%%%%%%

\subsection{Design rule 2}
\begin{figure}[htbp]
\centering
	\includegraphics [width = \columnwidth]{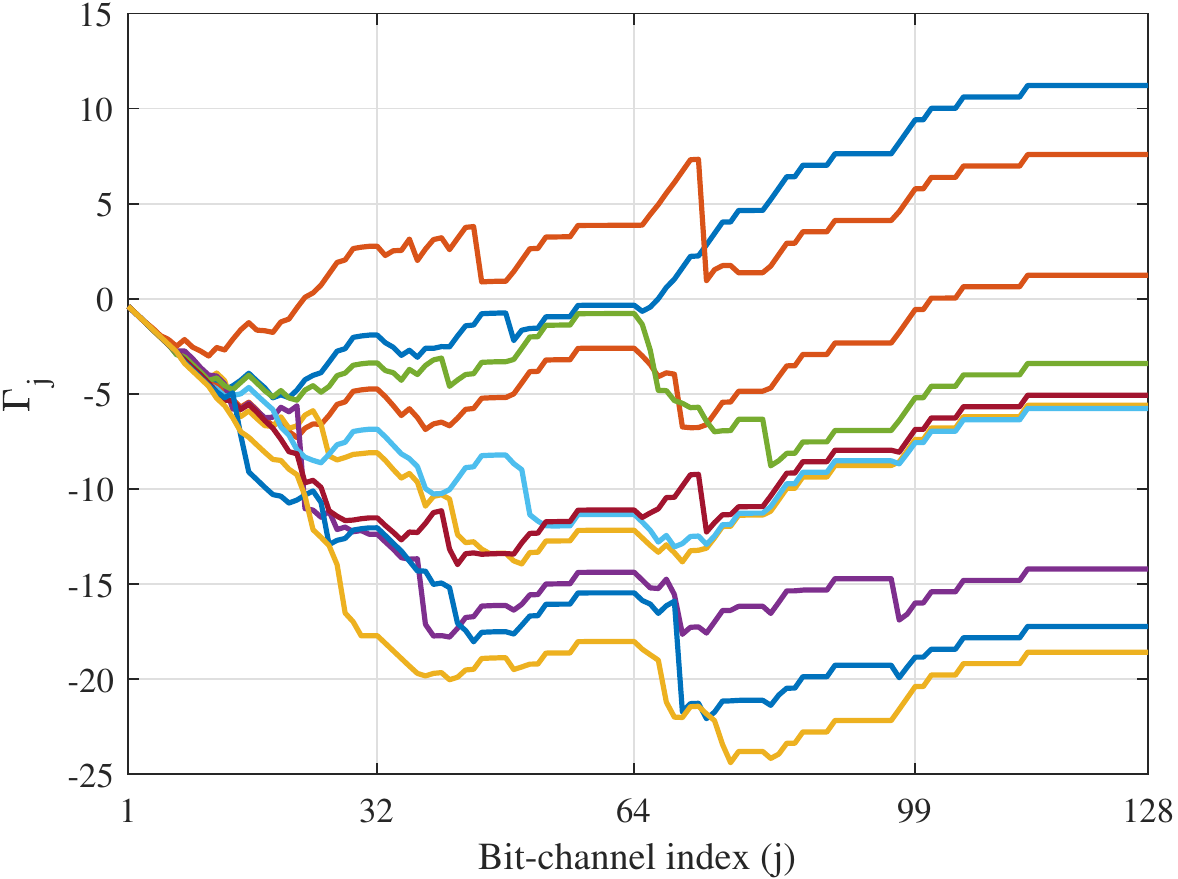}
	\caption{10 sample partial path metrics v. bit indices for K = 64 with $b_i^{(I)} = I(W_N^{(i)})$ and $b_i^{(F)} = 0.4$ at 2.5~dB SNR value.} 
	\label{fig: K64 fixedBF}
\end{figure}

In this part, without loss of generality, we will only discuss the $K = 64$ case.
For $K = 64$, using the bias values $b_i^{(I)} = I(W_N^{(i)})$ and $b_i^{(F)} = 0.4$, the PAC decoder performs as good as the decoder using rule 1 but with exponentially higher computation.
Since $b_i^{(F)} = 0.4 \nleq I(W_N^{(i)}) $, for frozen bits, the partial path metric becomes a negative number on average and results in an exponential growth in the computation of the sequential decoding. 
For the information bits, $b_i^{(I)} = I(W_N^{(i)}) $ and both inequalities of \ref{eq: av. ineq. bias} are satisfied. 
Hence, the metric growth value for the information bits will be zero on average. 
The partial path metric of design rule 2 is shown in Fig. \ref{fig: K64 fixedBF}. 

Computation wise, design rule 2 differs from design rule 1.
Rule 1 requires a high decoding effort for decoding the information bits.
In contrast, rule 2 requires a high decoding effort for decoding the frozen bits.
As we already mentioned, to have a low computation sequential decoding, the decoder should decode the first bits with low decoding effort.
Since for PAC codes, the first bits are generally frozen, rule 2 requires exponentially higher computation compared to rule 1.

\subsection{Design rule 3}
Design rule 3 is motivated by the fact that the sequential decoding is a greedy tree-search algorithm similar to the shortest path graph-search Dijkstra algorithm. 
% The greedy approach favors making a decision that is the best at present.
Dijkstra algorithm always finds the shortest path on a graph when the edges have positive weights and may or may not succeed in finding the shortest path when some edges have negative weight values.
With similar reasoning, both design rules mentioned in the previous subsections may or may not work with a reasonable amount of decoding computations for a given codeword. 
Both beforehand expressed rules present propitious error performance for the PAC codes of length 128 but required exponentially high computation.

\begin{figure}[htbp]
\centering
	\includegraphics [width = \columnwidth]{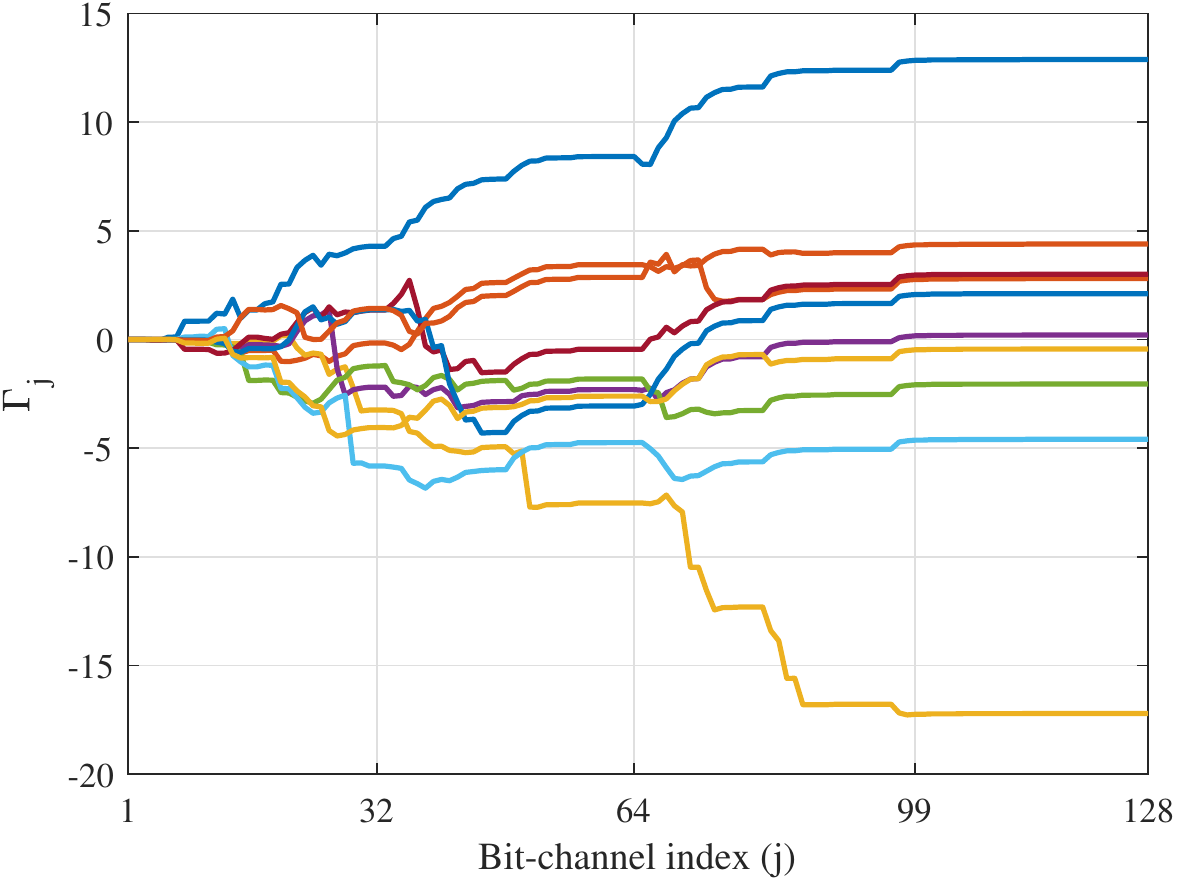}
	\caption{10 sample partial path metric v. bit indices for K = 64 with a bias value $I(W_N^{(i)})$ for all the bits at 2.5~dB SNR value.} 
	\label{fig: K64 bias IW}
\end{figure}

With choosing $b_i^{(I)}$ and $b_i^{(F)}$ both equal to the bit-channel mutual information $I(W_N^{(i)})$, both of (\ref{eq: av. ineq. bias}) inequalities are satisfied and the partial path metric of each branch becomes zero on average.
For all the bits, as $- b_i = - I(W_N^{(i)}) < 0$, rule 3 will prevent the decoder from going to a wrong path most of the times. 
As a fortiori, Fig. \ref{fig: K64 bias IW} demonstrates that using the design rule number 3, the average partial path metrics at each level is 0 in rule 3.
Hence, by employing design rule 3, the decoder will decode every bit with a lower computation than the design rule 1 and 2. 
Notice that in this figure, the plots are almost flat in the bits that are nearly error-free ($I(W_N^{(i)}) \approx 1$) or
completely noisy ($I(W_N^{(i)}) \approx 0$).

\begin{figure}[htbp]
\centering
	\includegraphics [width = \columnwidth]{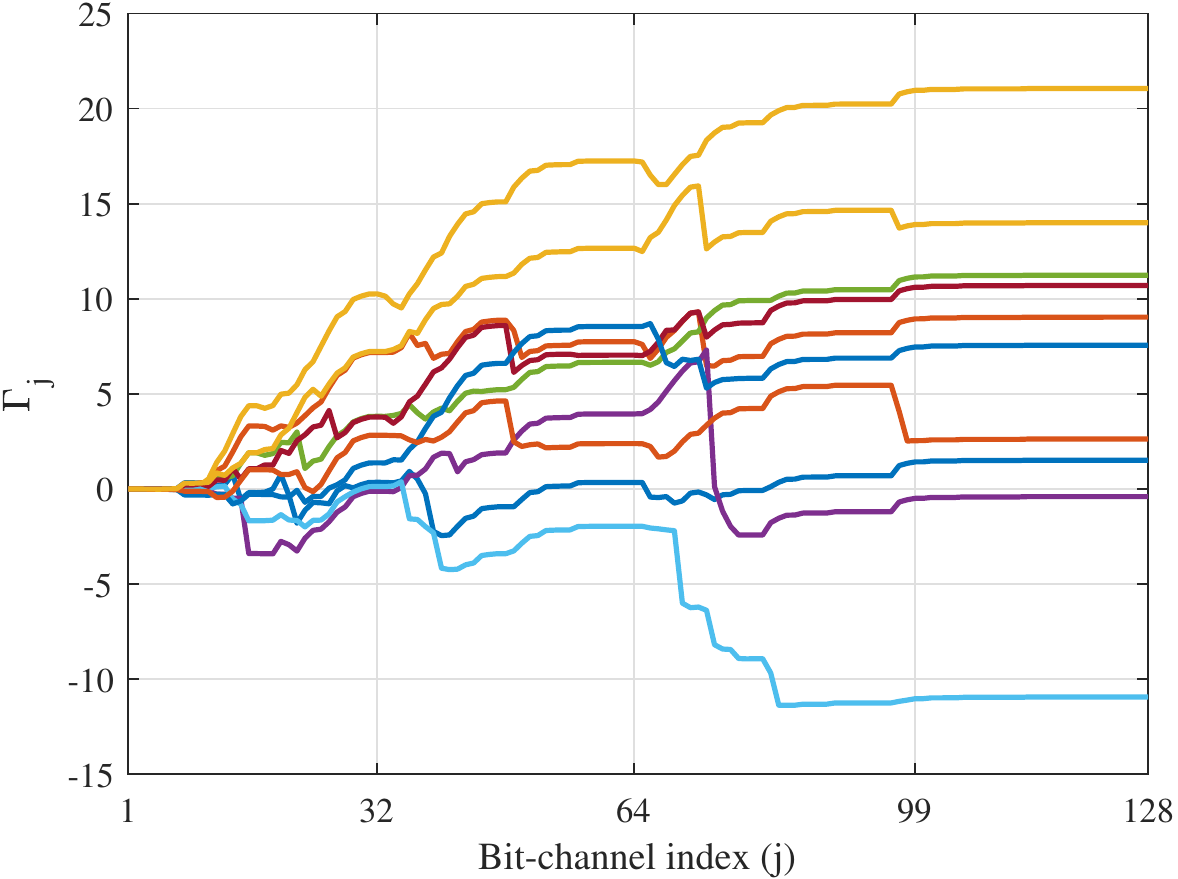}
	\caption{10 sample partial path metric v. bit indices for K = 64 with a bias value $E_0(1,W_N^{(i)})$ for all the bits at 2.5~dB SNR value.} 
	\label{fig: K64 bias E0}
\end{figure}

Alternatively, using $b_i = E_0(1, W_N^{(i)})$ for the bit-channel bias values satisfies both inequalities of (\ref{eq: av. ineq. bias}) strictly. 
Hence, for these bias values, the partial path metrics should become positive on average. Fig. \ref{fig: K64 bias E0} supports the latter statement. 
As a result, the computation with bias $b_i = E_0(1, W_N^{(i)})$ should be somewhat lower than with bias $b_i = I(W_N^{(i)})$ at a cost of negligible sacrifice of error performance.
The negligibility of error performance degradation and computation improvement is because of the polarization effect on both $E_0(1, W_N^{(i)})$ and $I(W_N^{(i)})$.
Note that in the conventional CCs, choosing the bias value greater than the cutoff rate and close to the capacity results in the exponential growth of computation. 
In this figure, the plots are almost flat in the nearly error-free bits or completely noisy bits, which supports that bit-channel cutoff rates are completely polarized in these bits.

%%%%%%%%%%%%%%%%%%%%%%%%%%%%%%%%%%%%%%%%%%%%%%%%%%%%%%%%%%%%%%%%%%%%%%%%%%%%%%%%%%%%%%%%%%%%%%%%%%%%%%%%%%%%%%%%%%%%%%%%%%%%%%%%%%%%%%

\section{Computation and Error-correction performance results} \label{sec: info_frozen simulation}
Considering \eqref{eq: av. ineq. bias}, only design rule 3 of the previous section satisfies both the inequalities. 
To understand the effect of the information and the frozen bits on error-correction performance and computation of the PAC codes separately, we split the design rule 3 into two cases based on the information bits and the frozen bits entitled item 1 and 2, respectively.
Finally, in item 3, we investigate the error-correction performance and computation results using design rule 3, which benefits from combining item 1 and item 2.

\subsection{Item 1: information bits}
\begin{figure}[t] 
\centering
	\includegraphics [width = \columnwidth]{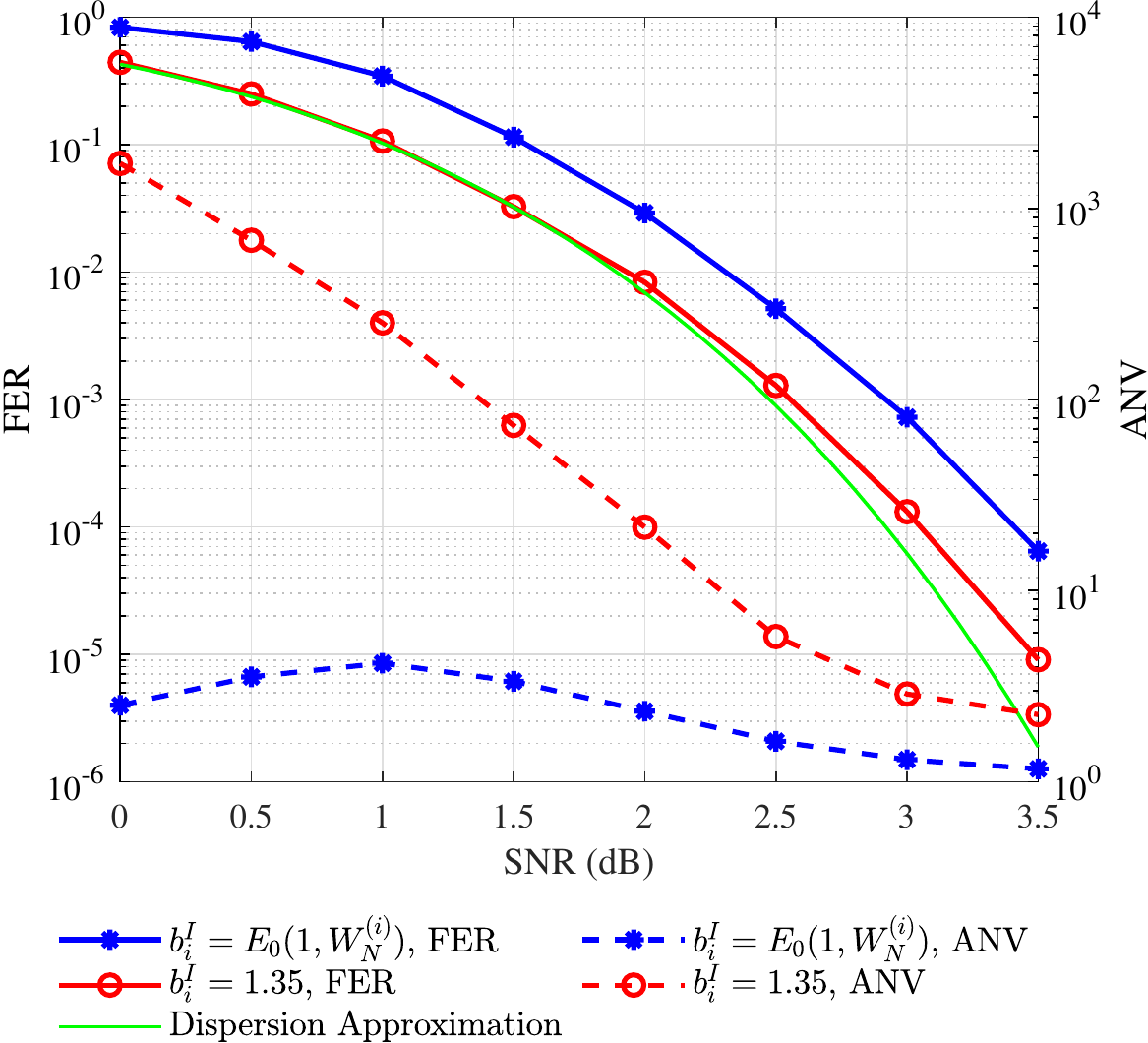}
	\caption{FER performance comparison of PAC codes with different bias values.} 
	\label{fig: FER PAC fixed and E0 (info)}
\end{figure}

As the study of the design rule three suggests, choosing the bit-channel bias values according to $E_0(1, W_N^{(i)})$ or $I(W_N^{(i)})$ can result in low computational decoding. 
In this subsection, we consider the bit-channel bias values of the information bits as $E_0(1, W_N^{(i)})$ and we fix the bit-channel bias values of the frozen bits to zero. 
Fig. \ref{fig: FER PAC fixed and E0 (info)} compares this choice with the PAC code using $b_i^{I} = 1.35$ and $b_i^{F} = 0$ bias values.
As we can observe from the figure, there is a significant improvement in the computation when using $b_i^{I} = E_0(1, W_N^{(i)})$ and $b_i^{F} = 0$ bias values.
Note that this well amount of decrease in the computation comes at the cost of sacrificing the PAC codes' error performance.
Dispersion approximation is also provided in this figure.

\begin{figure}[h] 
\centering
	\includegraphics [width = \columnwidth]{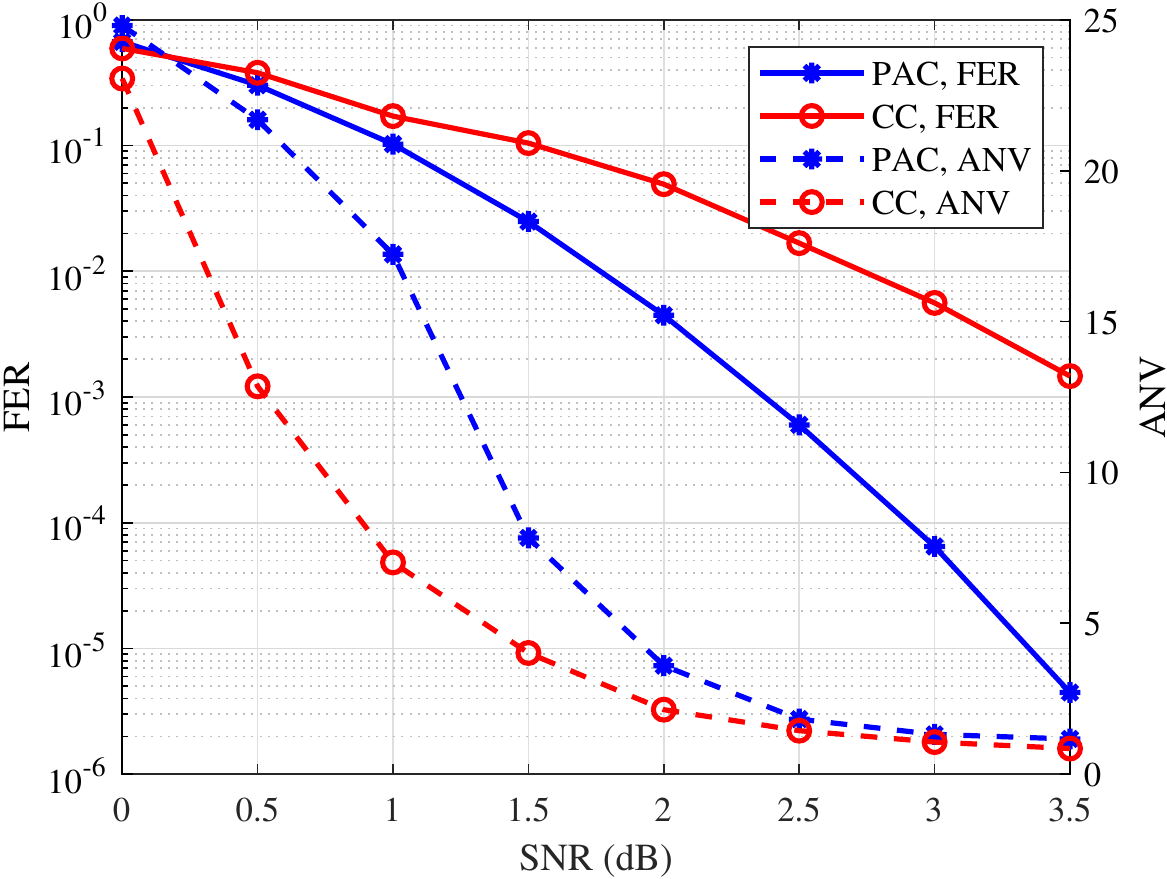}
	\caption{FER performance comparison of PAC and CC codes under sequential decoding.} 
	\label{fig: FER E0 and CC (info)}
\end{figure}

Moreover, Fig. \ref{fig: FER E0 and CC (info)} compares a PAC$(128,59)$ code using $b_i^{I} = E_0(1, W_N^{(i)})$ and $b_i^{F} = 0$ bias values with a CC$(140,64)$. 
As the figure shows, there is a significant improvement in the FER of the PAC code, whereas the ANV of the PAC sequential decoding is almost the same as the ANV of the CC sequential decoding. 
This improvement of the PAC code's error-correction performance is due to the polarization effect on the channel.
Sequential decoding for both PAC and CC codes have almost the same ANV result, and for both of them, the computation suffers from the cutoff rate phenomenon (ANV value for both of them increases for the SNR values less than $R_0 \approx 2.0$).
Note that to calculate the ANV for both codes, we count the number of visits per codeword and divide it by 128.

\subsection{Item 2: frozen bits}
\begin{figure}[htbp]
\centering
	\includegraphics [width = \columnwidth]{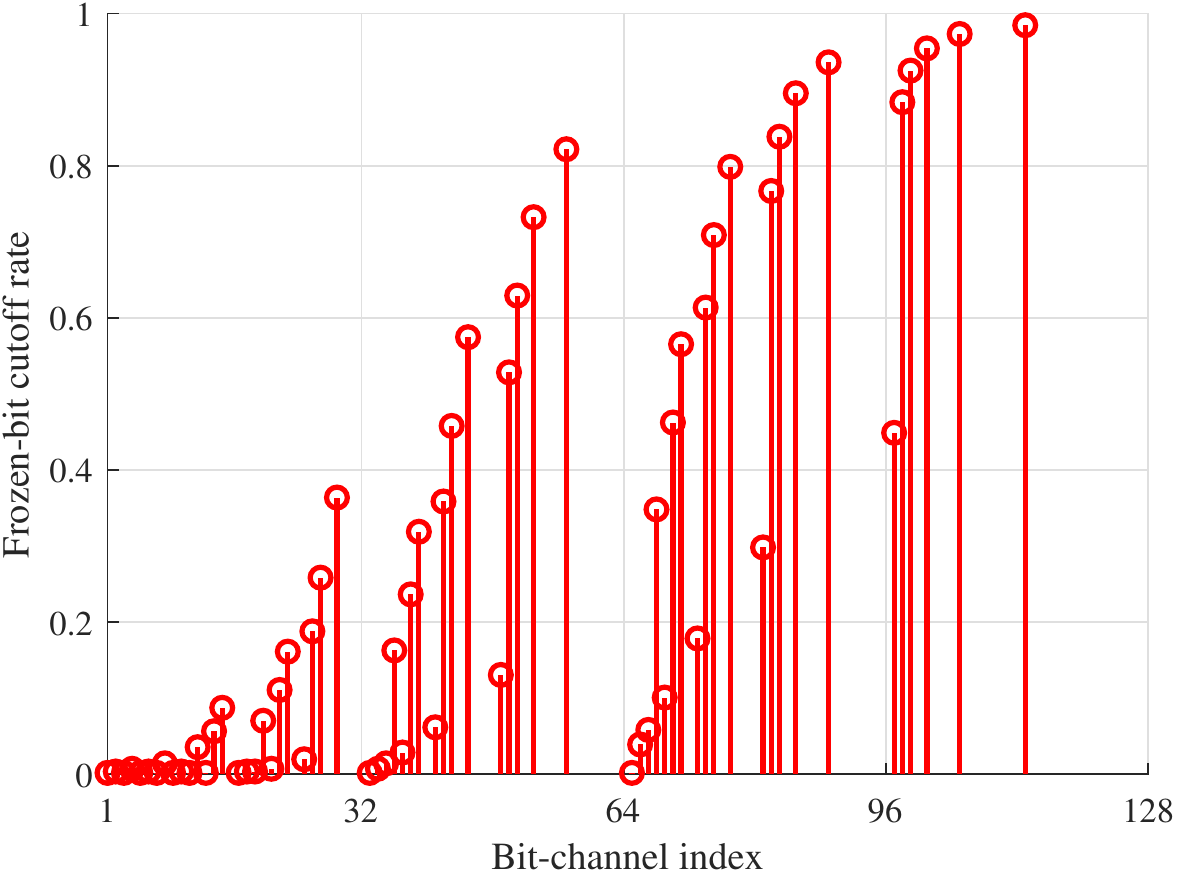}
	\caption{Polarized $E_0(1,W_N^{(i)})$ values of the frozen bits at 2.5~dB SNR value.} 
	\label{fig: frozen bit channel E0}
\end{figure}

This subsection studies the effect of the frozen bits on the error-correction performance and computation of PAC codes by using $b_i^{F} = E_0(1, W_N^{(i)})$ and $b_i^{I} = 0$ bias values. 
Note that these choices of bias values satisfy \eqref{eq: av. ineq. bias}. 
The idea of using frozen bits for decoding PAC codes is motivated by the fact that for the short-length codes, the bit channels are not entirely polarized. 
Moreover, when using the RM scoring rule in constructing PAC codes, the $E_0(1, W_N^{(i)})$ of the frozen bits are not absolute zero as Fig. \ref{fig: frozen bit channel E0} illustrates. 

\begin{figure}[h] 
\centering
	\includegraphics [width = \columnwidth]{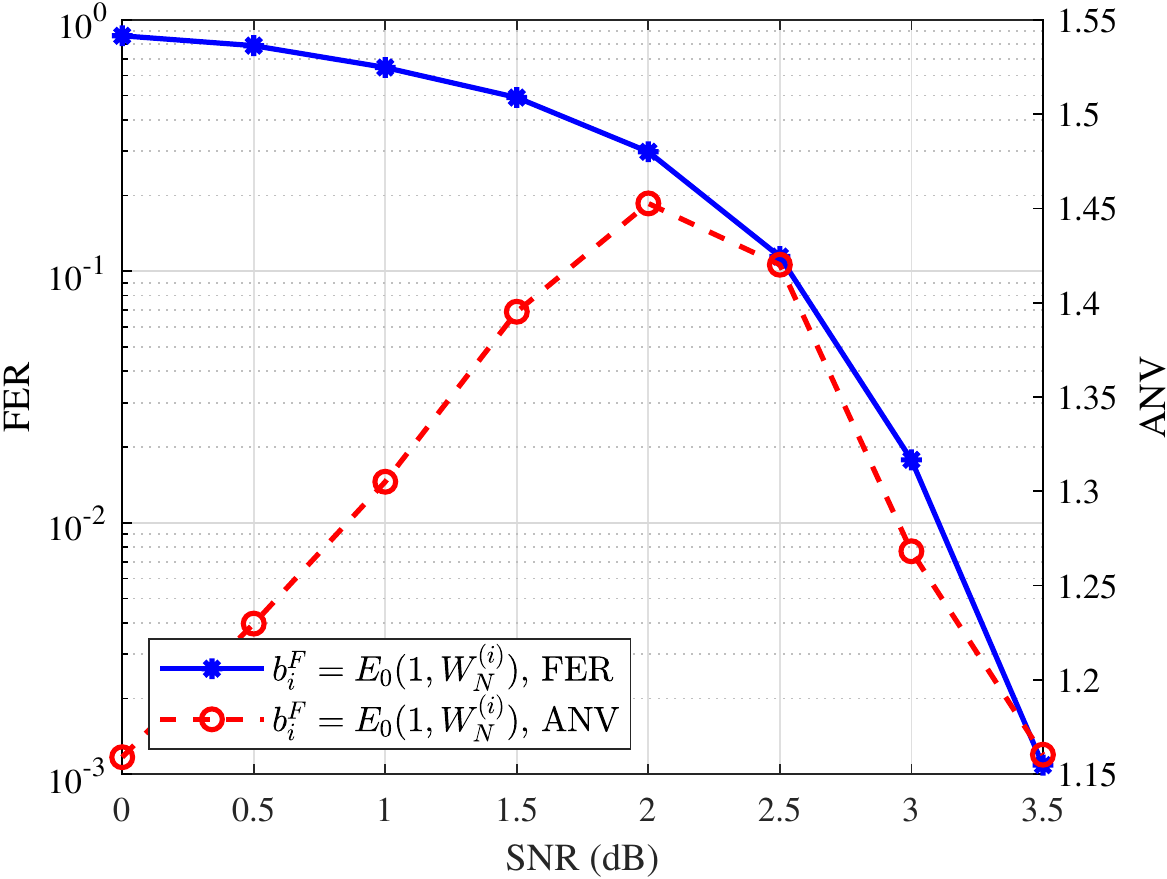}
	\caption{FER performance of PAC when bias $E_0(1,W_N^{(i)})$ is used just for the frozen bits.} 
	\label{fig: FER frozen}
\end{figure}

Fig. \ref{fig: FER frozen} plots the error-correction performance and computation of a PAC code employing $b_i^{F} = E_0(1, W_N^{(i)})$ and $b_i^{I} = 0$. 
As expected, since both inequalities of \eqref{eq: av. ineq. bias} are satisfied; the ANV is less than 1.5 on average. 

Recall that the crucial attribute of sequential decoding is that if the wrong turns can be discovered early enough, the saving in the number of computations (measured in terms of the number of branches explored) would be exponential. 
Since in the PAC codes the first bits are generally frozen bits, the choice of $b_i^{F} = E_0(1, W_N^{(i)})$ can result in an exponential reduction in computation.


%%%%%%%%%%%%%%%%%%%%%%%%%%%%%%%%%%%%%%%%%%%%%%%%%%%%%%%%%%%%%%%%%%%%%%%%%%%%%%%%%%%%%%%%%%%%%%%%%%%%%%%%%%%%%%%%%%%%%%%%%%%%%%%%%%%%%%

\subsection{Item 3}\label{sec: optimal IW simulation}
In the sequential decoding of the conventional CCs, because of the gap between the cutoff rate and the channel capacity, choosing the metric function bias value close to the channel capacity will result in an exponential growth in the computation.
However, for PAC codes, because of the small gap between the bit-channel cutoff rates $E_0(1, W_N^{(i)})$ and bit-channel capacities $I(W_N^{(i)})$, there is a negligible difference in the decoding computation and error-correction performance while using either of them as the bias value.

\begin{figure}[h] 
\centering
	\includegraphics [width = \columnwidth]{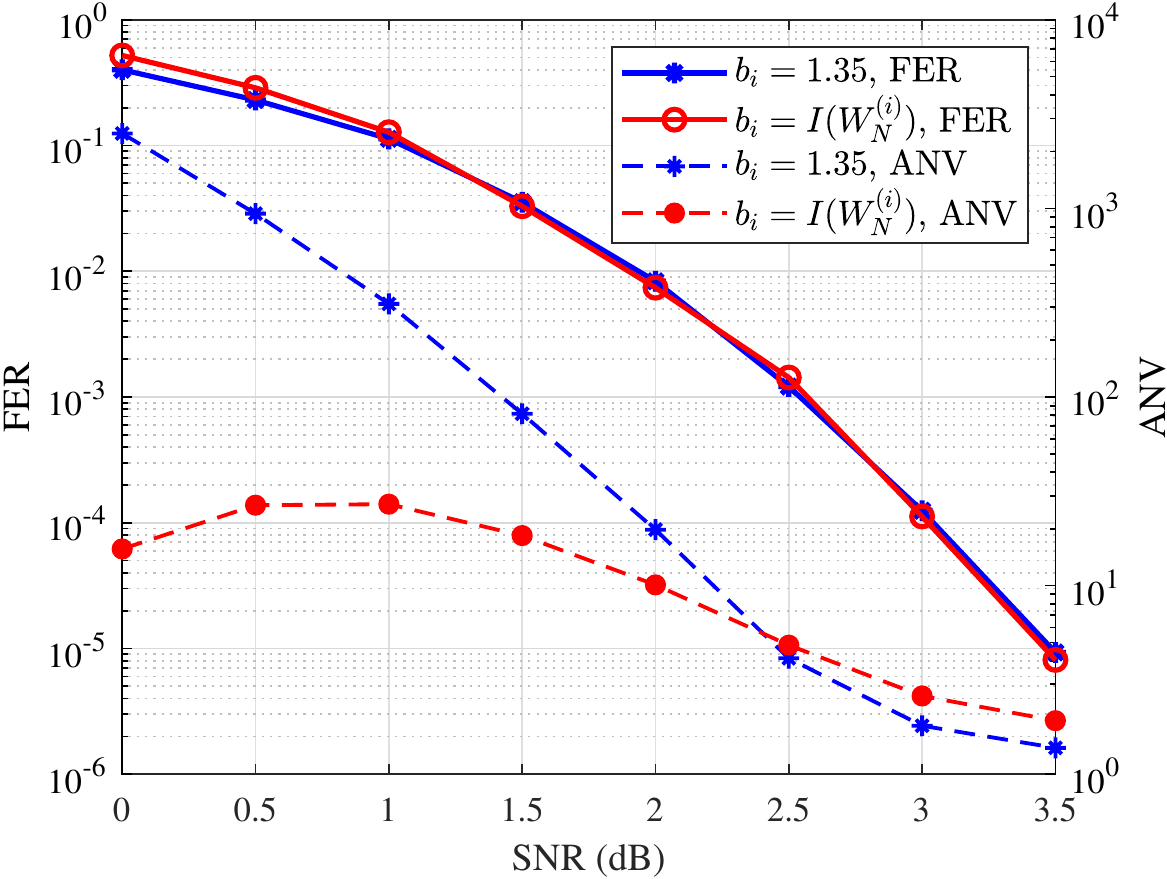}
	\caption{Comparing FER performance of the PAC codes.} 
	\label{fig: FER IW (all) and 1.35}
\end{figure}

Fig. \ref{fig: FER IW (all) and 1.35} plots a comparison of FER and ANV of  a PAC$(128,64)$ code using the bias values $b_i = I(W_N^{(i)})$ versus the bias values $b_i^{I} = 1.35$ and $b_i^{F} = 0$. 
From this figure, we can conclude that both choices of bias values result in the same error-correction performance with $b_i = I(W_N^{(i)})$ having exponentially smaller computation.

\begin{figure}[t] 
\centering
	\includegraphics [width = \columnwidth]{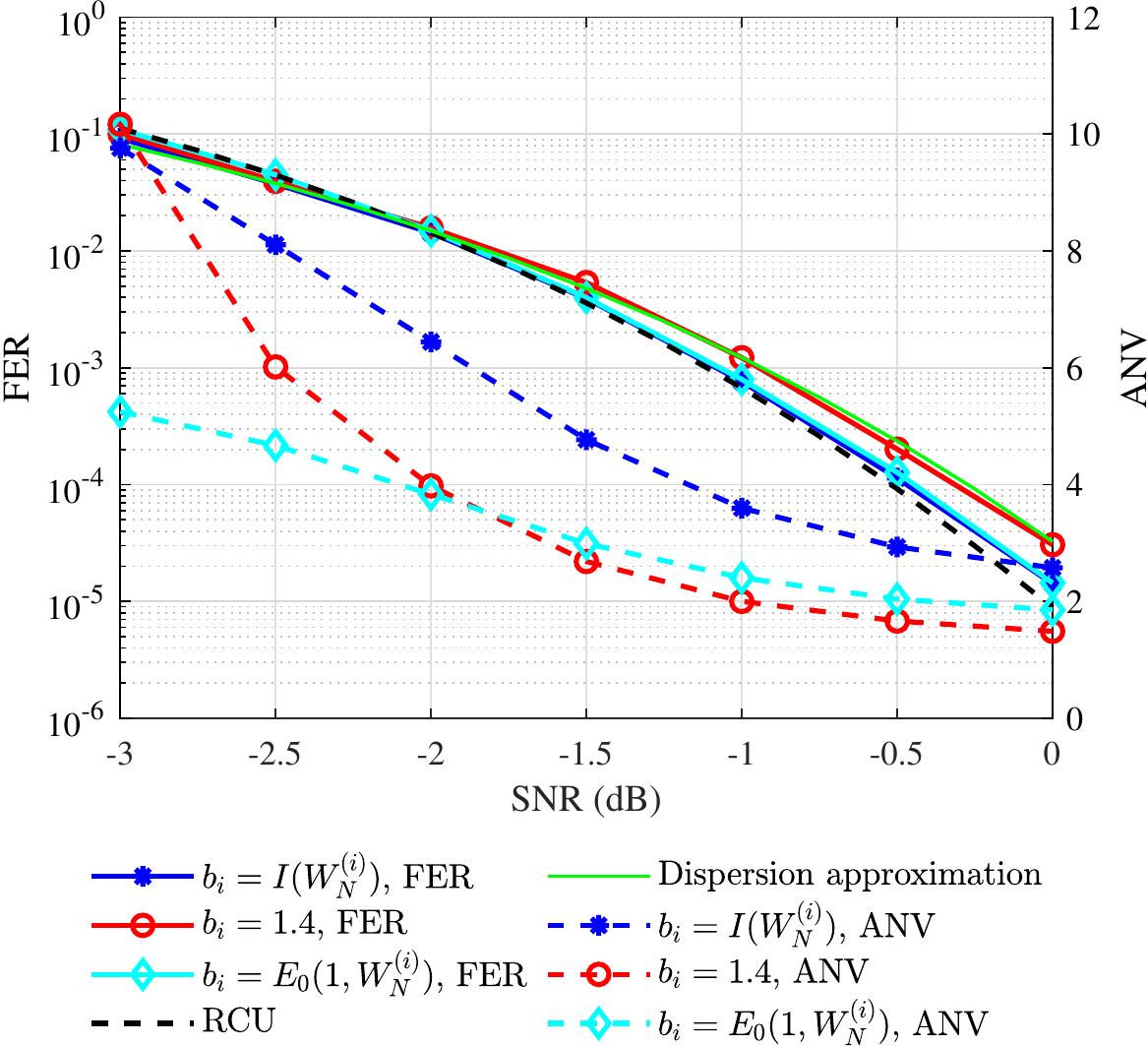}
	\caption{FER performance comparison of PAC codes for K = 29.} 
	\label{fig: K29 IW (all) and 1.4}
\end{figure}

Moreover, Fig. \ref{fig: K29 IW (all) and 1.4} plots a comparison of FER and ANV of a PAC$(128,29)$ code using the bit-channel bias values $b_i = I(W_N^{(i)})$ and $b_i = E_0(1, W_N^{(i)})$ versus using the fix bias values $b_i^{I} = 1.4$ and $b_i^{F} = 0$. 
From this figure, we can conclude that choosing $b_i = I(W_N^{(i)})$ or $b_i = E_0(1, W_N^{(i)})$ results in better error-correction performance and comparable computation compared  to the fix bias values $b_i^{I} = 1.4$ and $b_i^{F} = 0$.
The random-coding union (RCU) bound and dispersion approximation are also provided in this figure \cite{polyanskiy2010channel}.

\begin{figure}[t] 
\centering
	\includegraphics [width = \columnwidth]{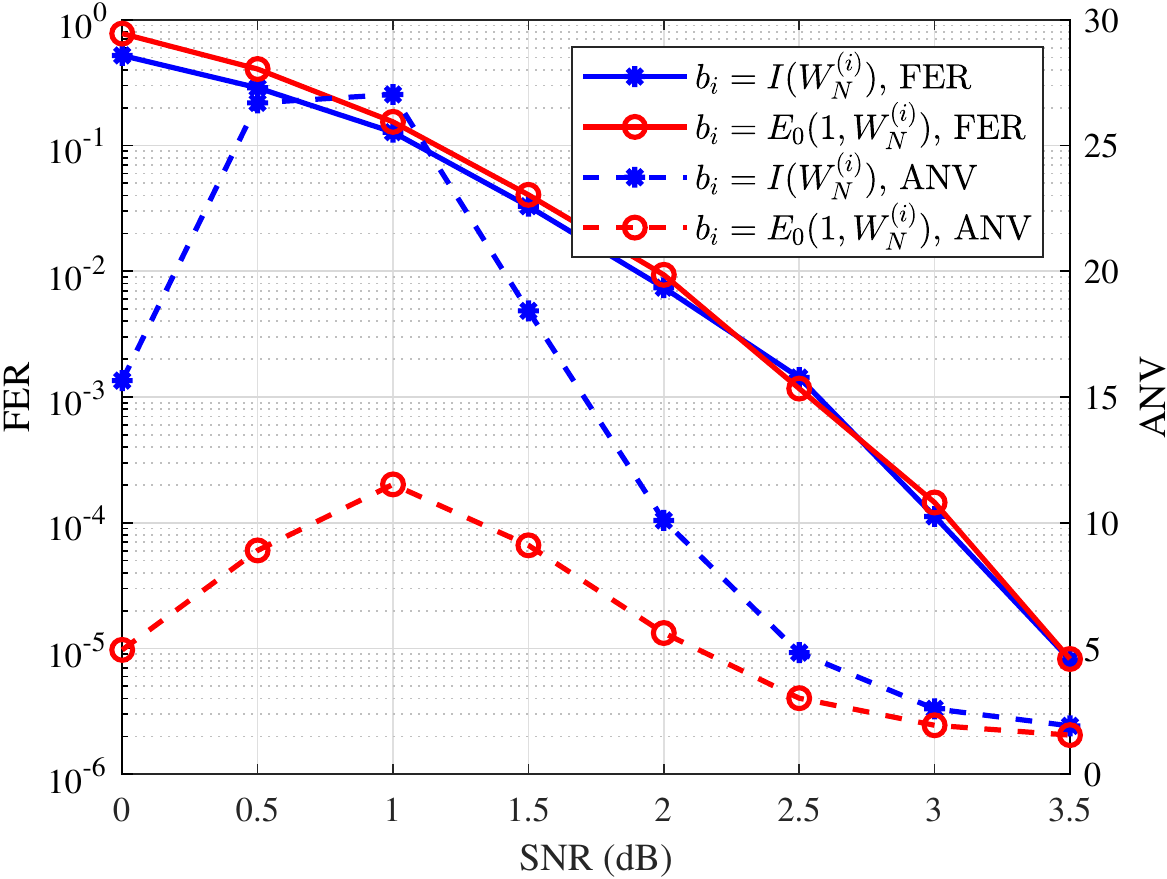}
	\caption{FER performance of PAC codes with bias $I(W_N^{(i)})$ and $E_0(1,W_N^{(i)})$.} 
	\label{fig: FER IW and E0}
\end{figure}

Fig. \ref{fig: FER IW and E0} compares FER and ANV of PAC$(128,64)$ code using the bias values $b_i = I(W_N^{(i)})$ versus $b_i = E_0(1, W_N^{(i)})$.
This figure strengthens our previous idea that since $E_0(1, W_N^{(i)})$ is slightly lower than $I(W_N^{(i)})$, the computation due to using $b_i = I(W_N^{(i)})$ would be higher. 
The performance improvement marginally occurs only at the low SNR values (noisy channels). 

%%%%%%%%%%%%%%%%%%%%%%%%%%%%%%%%%%%%%%%%%%%%%%%%%%%%%%%%%%%%%%%%%%%%%%%%%%%%%%%%%%%%%%%%%%%%%%%%%%%%%%%%%%%%%%%%%%%%%%%%%%%%%%%%%%%%%%

\section{Scaling bias values}\label{sec: scaling bias}
In this section, we investigate the behavior of the error-correction performance and the computation of search-limited and search-unlimited PAC decoder using scaled bias values as $\alpha \times I(W_N^{(i)})$ or $\alpha \times E_0(1, W_N^{(i)})$, where $\alpha$ is a constant value. 
We will show that for a search-limited sequential decoder to achieve a better error-correction performance, the bias value is required to be scaled down.

\subsection{Search-unlimited PAC codes}
\begin{figure}[htbp]
\centering
	\includegraphics [width = \columnwidth]{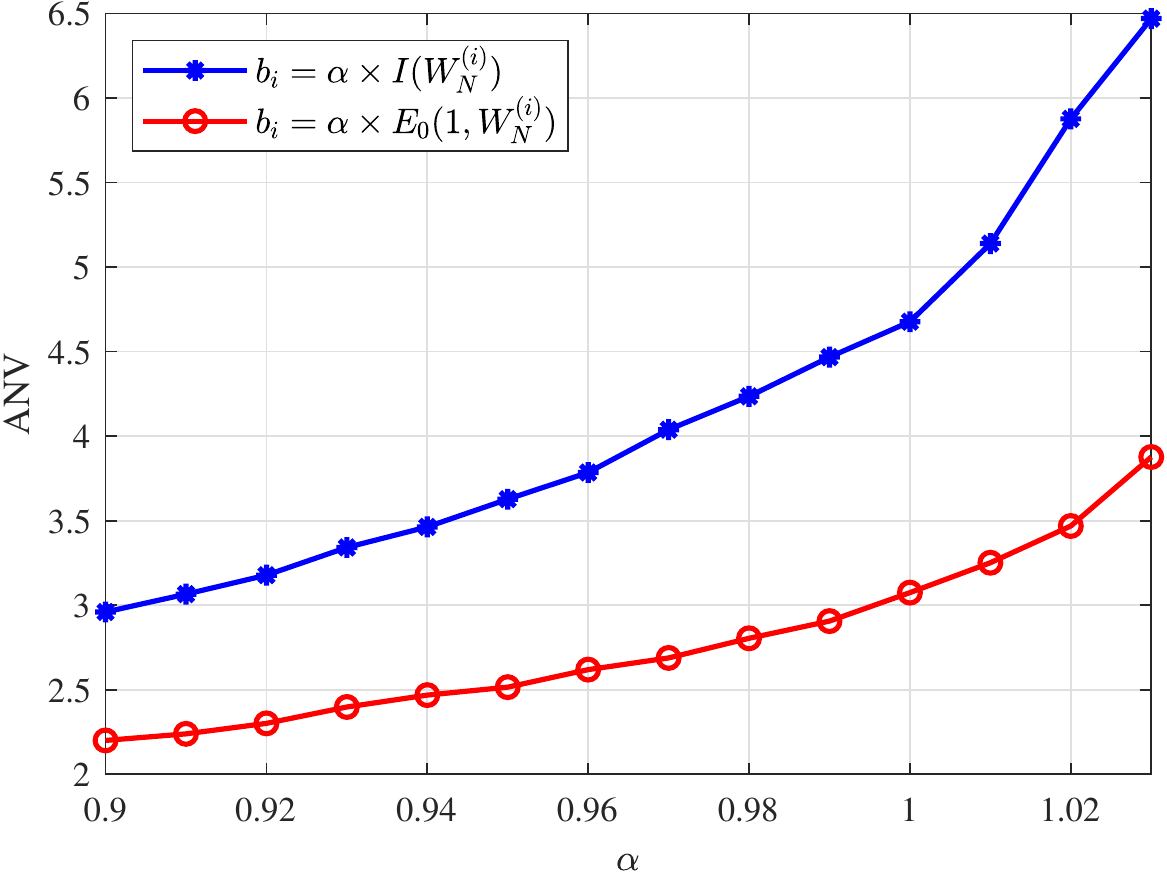}
	\caption{ANV v. $\alpha$ for a (128,64) PAC code with an unlimited-search Fano decoding at 2.5 dB SNR value, delta = 2.} 
	\label{fig: ANV alphaIW unlimited}
\end{figure}

In this part, we investigate the effect of bias scaling on the ANV and FER performance of the PAC codes under no computation limit.
Fig. \ref{fig: ANV alphaIW unlimited} plots ANV of the PAC decoder using bias of $\alpha \times I(W_N^{(i)})$ and $\alpha \times E_0(1, W_N^{(i)})$ for various values of $\alpha$. 
As shown in this figure, as $\alpha$ increases, the ANV increases as well for both choices of bias values.
Note that, for the choice of the bias value as $\alpha \times I(W_N^{(i)})$, the slope of the ANV increases after $\alpha = 1$, whereas for $\alpha \times E_0(1, W_N^{(i)})$, the increase occurs after $\alpha = 1.02$.

\begin{figure}[htbp]
\centering
	\includegraphics [width = \columnwidth]{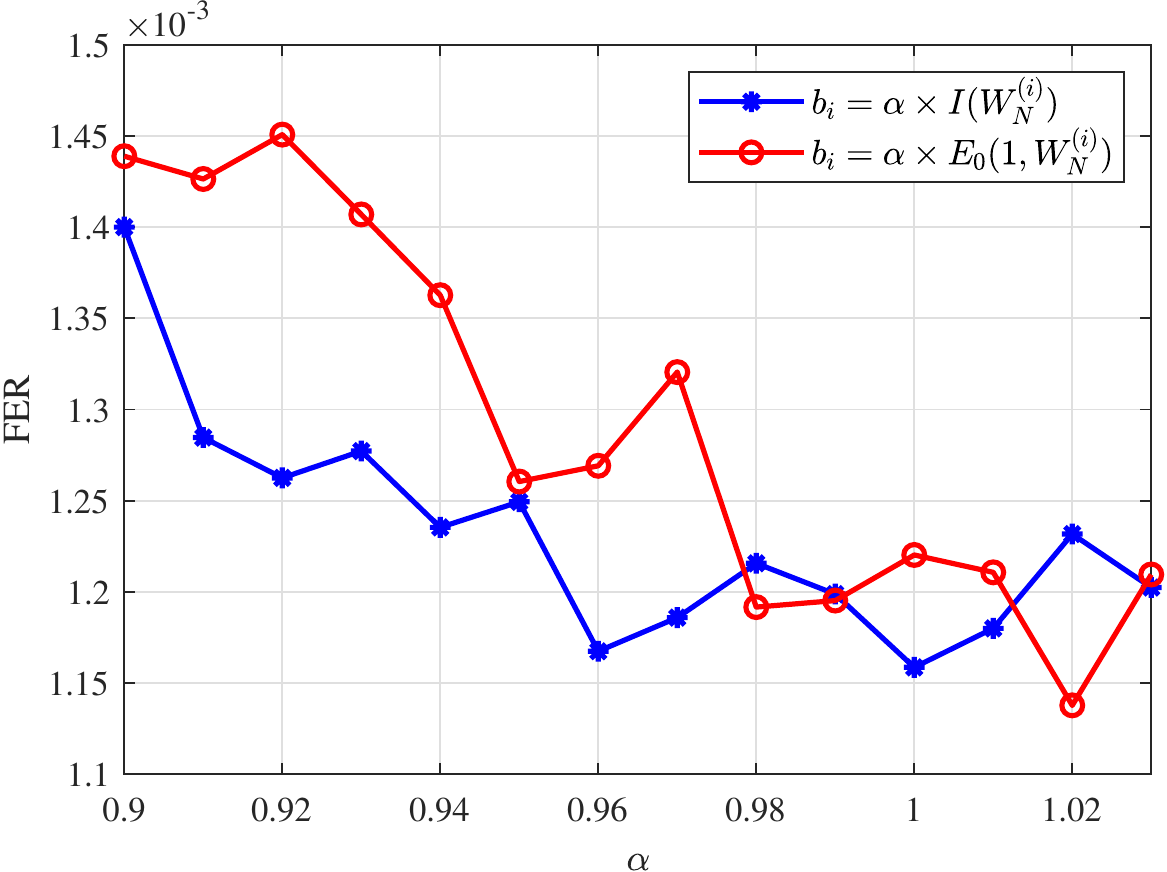}
	\caption{FER v. $\alpha$ for a (128,64) PAC code with an unlimited-search Fano decoding at 2.5 dB SNR value, delta = 2.} 
	\label{fig: FER alphaIW unlimited}
\end{figure}

Fig. \ref{fig: FER alphaIW unlimited} plots the FER performance of the PAC decoders of Fig. \ref{fig: ANV alphaIW unlimited}.
From this figure, we notice that the best FER is achieved with the $\alpha$ values close to 1 and 1.02 for $b_i = \alpha \times I(W_N^{(i)})$ and $b_i = \alpha \times E_0(1, W_N^{(i)})$, and as $\alpha$ diverges from 1 and 1.02, the FER drops, respectively. 
Simulation results recommend using $1.02\times E_0(1, W_N^{(i)})$ instead of $1\times I(W_N^{(i)})$ to achieve a lower ANV with almost the same error-correction performance. 

\begin{figure}[htbp]
\centering
	\includegraphics [width = \columnwidth]{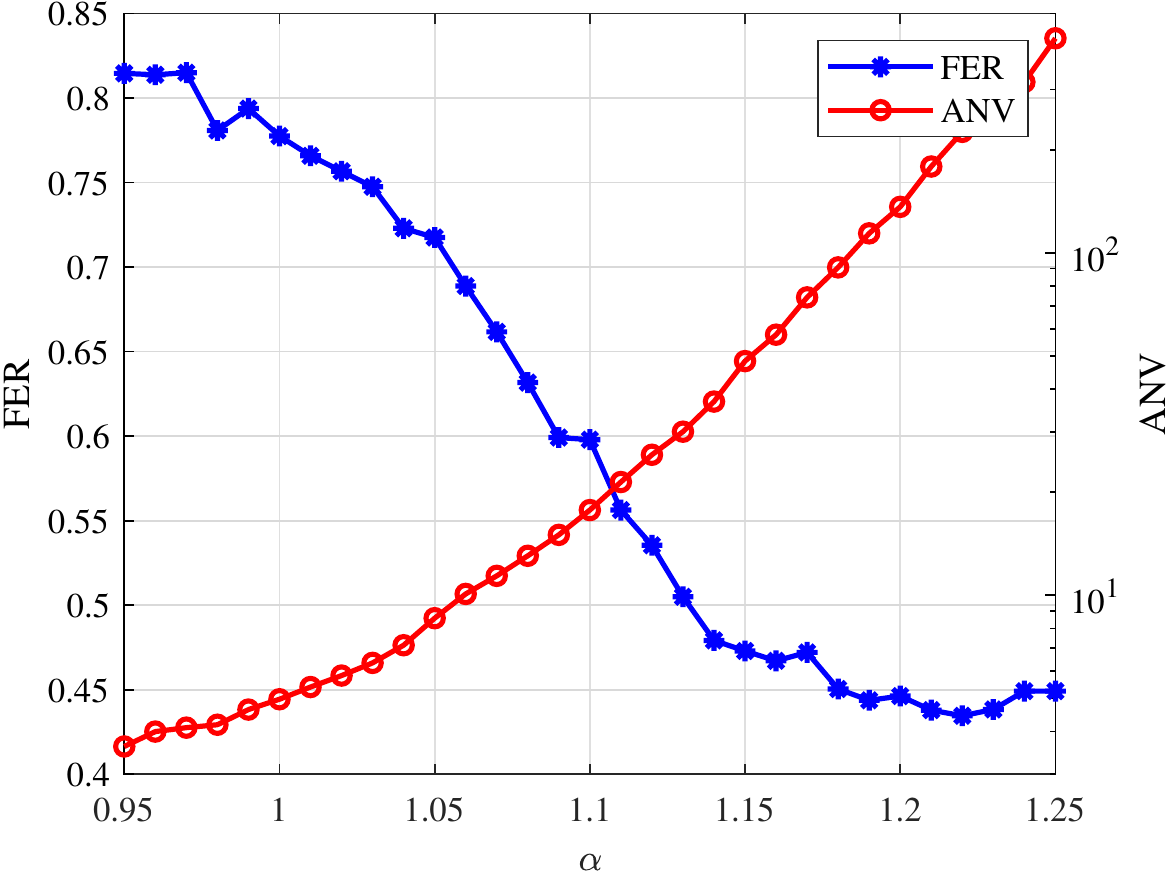}
	\caption{FER v. $\alpha$ for a (128,64) PAC code with an unlimited-search Fano decoding at 0 dB SNR value (noisy channel), delta = 2.} 
	\label{fig: FER_ANV E0 noisy_channel}
\end{figure}

Fig. \ref{fig: FER_ANV E0 noisy_channel} plots the FER and ANV versus $\alpha$ for the bias values as the bit-channel cutoff rates for a noisy channel (SNR = 0.0 dB). 
From Fig. \ref{fig: FER IW and E0} we can observe that there is a gap between the FER performance of using $b_i = E_0(1,W_N^{(i)})$ and the dispersion approximation.
To reduce this gap, Fig. \ref{fig: FER_ANV E0 noisy_channel} suggest increasing $\alpha$ value, but as $\alpha$ increases, the computation grows exponentially.
To summarize, the FER performance of PAC codes using $b_i = \alpha \times E_0(1,W_N^{(i)})$ can near the dispersion approximation at noisy channels at a cost of extreme computation increase.

In the next section, we will show that the effect of bias scaling on ANV and FER performance is more significant for the search-limited PAC decoder.

\subsection{Search-limited PAC codes}
This subsection investigates the effect of bias scaling on the error-correction performance and ANV of the search-limited PAC Fano decoder.
Throughout this paper, we use MNV $= 2^{12}$ for the search-limited PAC decoder.

\begin{figure}[htbp]
\centering
	\includegraphics [width = \columnwidth]{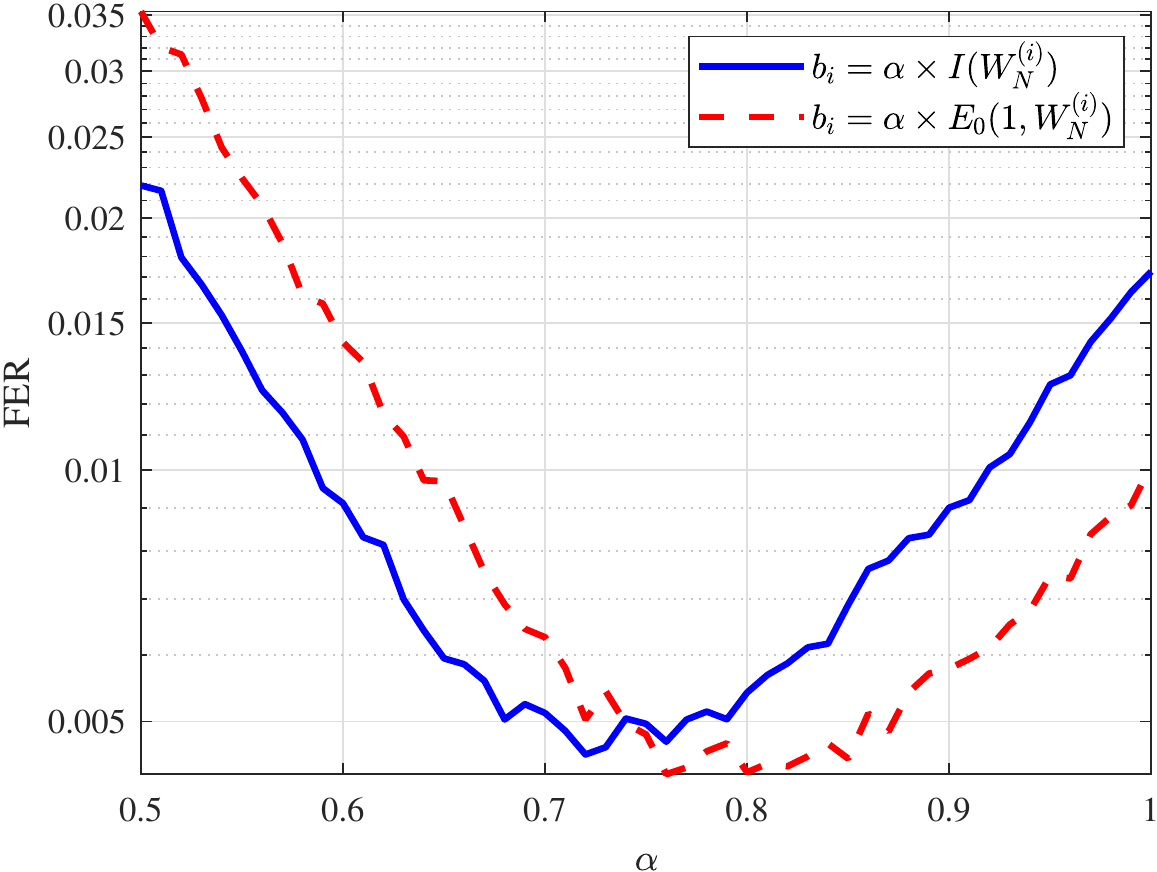}
	\caption{FER v. $\alpha$ for an $(128,64)$ PAC code with $\alpha \times E_0(1,W_N^{(i)})$ and $\alpha \times I(W_N^{(i)})$ bias values, MNV = $2^{12}$ at 2.5~dB SNR value, and threshold spacing $\Delta = 2$.} 
	\label{fig: limited FER alphaE0}
\end{figure}

Fig. \ref{fig: limited FER alphaE0} plots FER versus $\alpha$ for bias values of $\alpha\times I(W_N^{(i)})$ and $\alpha\times E_0(1,W_N^{(i)})$.
As we can see from this figure, using $\alpha = 0.72 $ and $\alpha = 0.76 $ result in the best FER values for $\alpha\times I(W_N^{(i)})$ and $\alpha\times E_0(1,W_N^{(i)})$ bias values, respectively.  
In our simulations (not reported here) we realised that when using the bias value equal to the $\alpha\times I(W_N^{(i)})$ or $\alpha\times E_0(1,W_N^{(i)})$ for $0 \leq \alpha \leq 1$, for large MNV values, choosing $\alpha$ value close to 1 results in better FER performance, whereas for smaller MNV values, a good FER performance can be obtained by choosing smaller $\alpha$ values.

\begin{figure}[htbp]
\centering
	\includegraphics [width = \columnwidth]{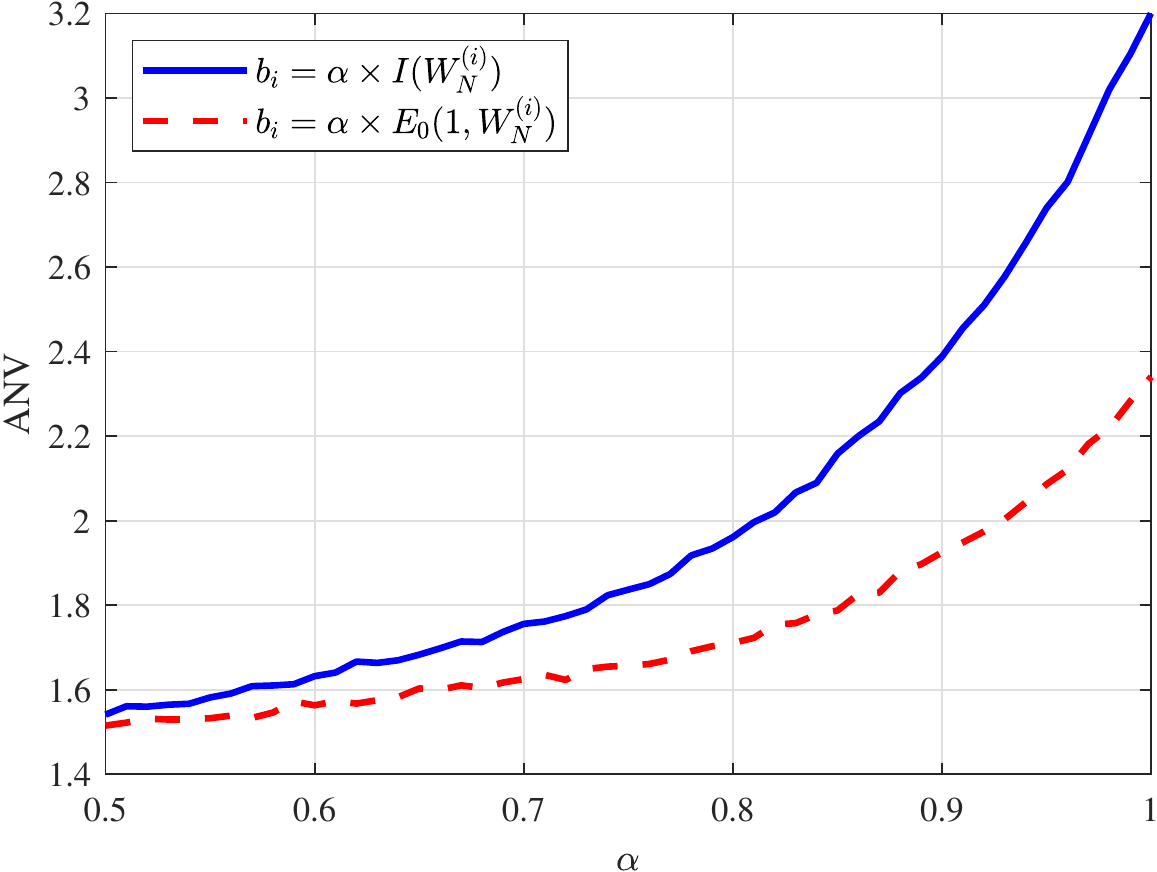}
	\caption{ANV v. $\alpha$ for an $(128,64)$ PAC code with $\alpha \times E_0(1,W_N^{(i)})$ and $\alpha \times I(W_N^{(i)})$ bias values, MNV = $2^{12}$ at 2.5~dB SNR value, and threshold spacing $\Delta = 2$.} 
	\label{fig: limited ANV alphaE0}
\end{figure}

Fig. \ref{fig: limited ANV alphaE0} plots ANV versus $\alpha$ for bias values of $\alpha\times I(W_N^{(i)})$ and $\alpha\times E_0(1,W_N^{(i)})$.
In this figure, we can observe that as $\alpha$ decreases, the ANV decreases as well.

%%%%%%%%%%%%%%%%%%%%%%%%%%%%%%%%%%%%%%%%%%%%%%%%%%%%%%%%%%%%%%%%%%%%%%%%%%%%%%%%%%%%%%%%%%%%%%%%%%%%%%%%%%%%%%%%%%%%%%%%%%%%%%%%%%%%%%

\section{Threshold spacing}\label{sec: threshold spacing}
In this section, we study the impact of the threshold spacing $\Delta$ on the error performance and computation of the PAC codes.
Different values of $\Delta$ will result in different FER and ANV values for both search-limited and -unlimited sequential decoding of the PAC codes. 
Based on the ANV and FER trade-off, deciding on the most suitable threshold spacing value is a matter of great concern.

\subsection{Search unlimited}
\begin{figure}[htbp]
\centering
	\includegraphics [width = \columnwidth]{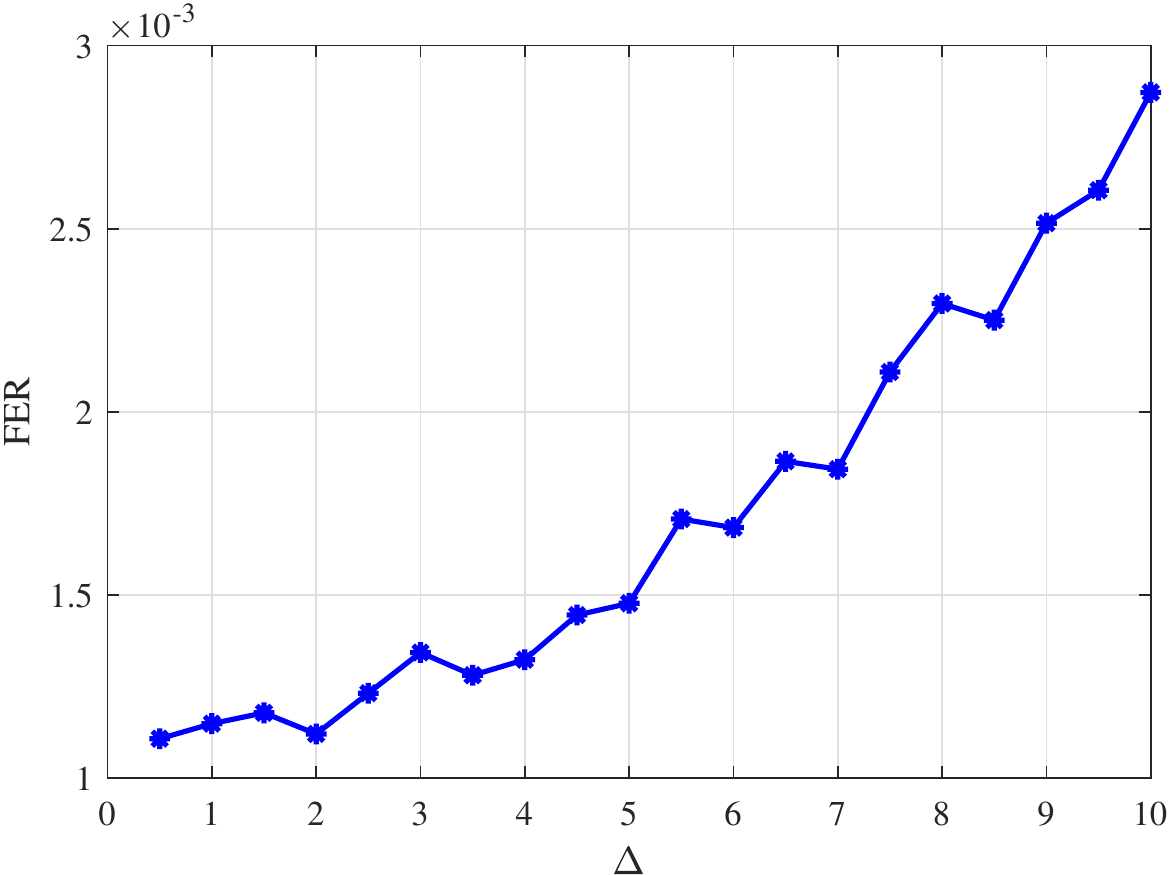}
	\caption{FER v. threshold spacing $\Delta$.} 
	\label{fig: FER v. delta}
\end{figure}

\begin{figure}[htbp]
\centering
	\includegraphics [width = \columnwidth]{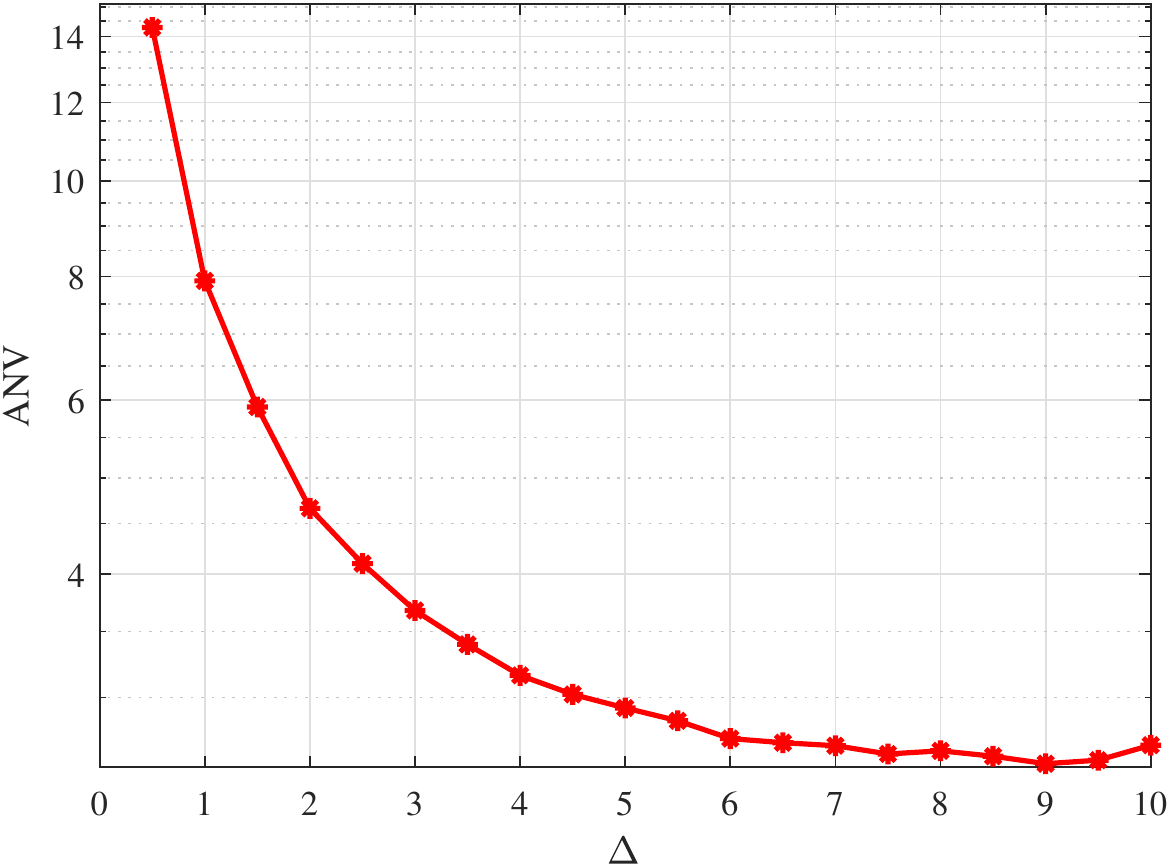}
	\caption{ANV  v. threshold spacing $\Delta$.} 
	\label{fig: ANV v. delta}
\end{figure}
Fig. \ref{fig: FER v. delta} shows the effect of $\Delta$ value on the error performance of a PAC(128,64) code at $2.5$~dB SNR value, where the bit-channel bias values are set to $I(W_N^{(i)})$.
As the figure displays, the FER performance degrades as the $\Delta$ value increases;
$\Delta = 0.5$ and $\Delta = 2$ result in the best FER, and $\Delta = 10$ results in the worst FER for this range of threshold spacing.
The corresponding ANV values are plotted in Fig. \ref{fig: ANV v. delta}. 
As the figure exhibits, the ANV value decreases by increasing the $\Delta$ value.
Combining Fig. \ref{fig: FER v. delta} and \ref{fig: ANV v. delta}, we can conclude that $\Delta =2$ is a reasonable choice for threshold spacing value to maintain a good trade-off between the FER performance and computation.

%%%%%%%%%%%%%%%%%%%%%%%%%%%%%%%%%%%%%%%%%%%%%%%%%%%%%%%%%%%%%%%%%%%%%%%%%%%%%%%%%%%%%%%%%%%%%%

\subsection{Search limited}
\begin{figure}[htbp]
\centering
	\includegraphics [width = \columnwidth]{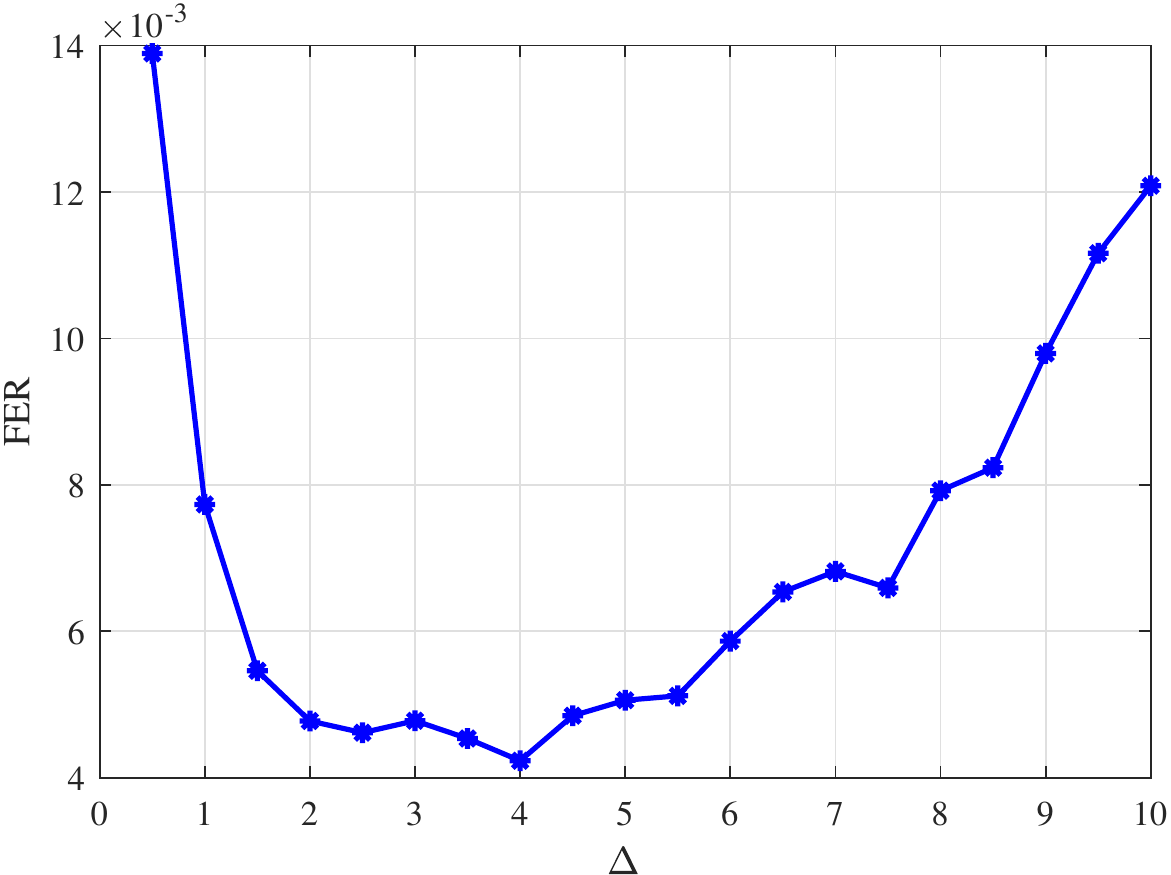}
	\caption{FER v. threshold spacing $\Delta$ parameter for a search limited (128,64) PAC code.} 
	\label{fig: limited FER v. delta}
\end{figure}

\begin{figure}[htbp]
\centering
	\includegraphics [width = \columnwidth]{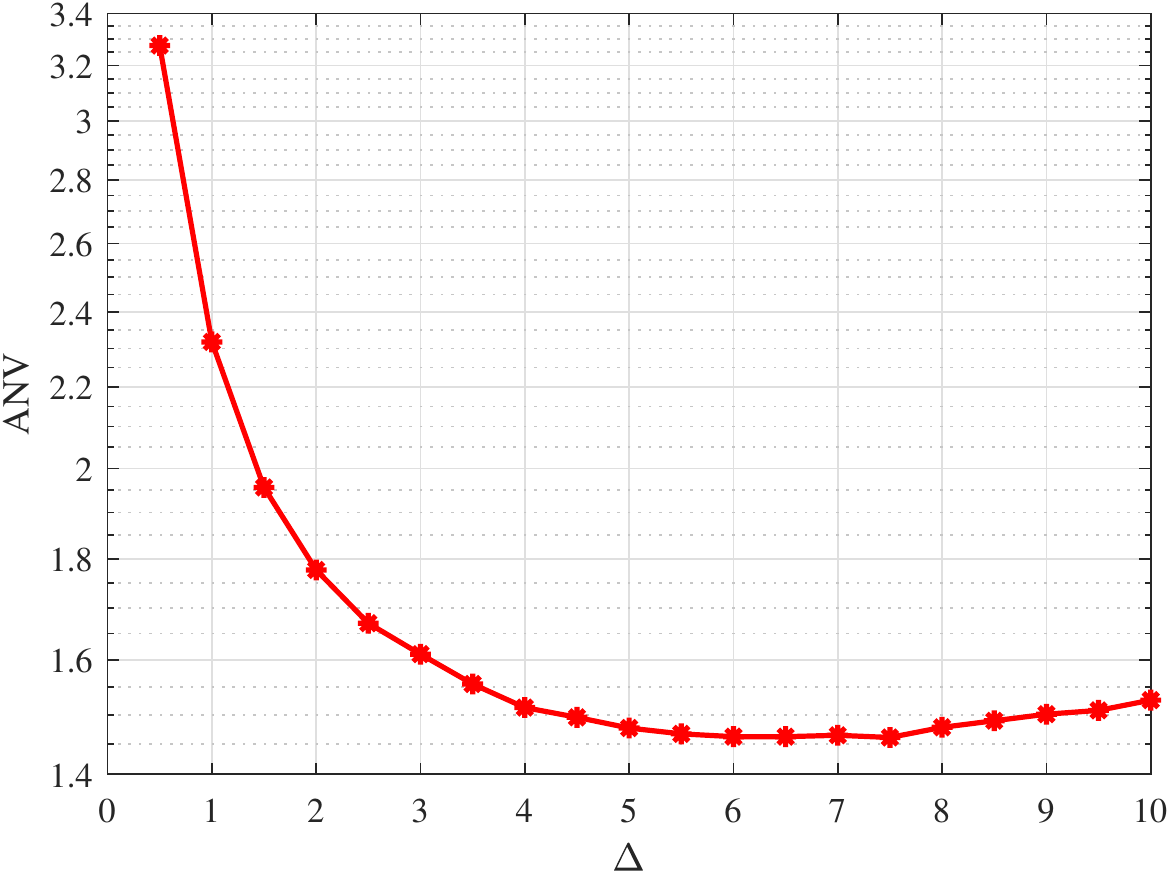}
	\caption{ANV v. threshold spacing $\Delta$ parameter for a search limited (128,64) PAC code.} 
	\label{fig: limited ANV v. delta}
\end{figure}
Search-limited PAC$(128,64)$ code with MNV $ = 2^{12}$ and bit-channel bias values equal to $0.72\times I(W_N^{(i)})$ at $2.5$~dB SNR is considered in this subsection.
The error performance results of the search-limited PAC code using different values of threshold spacing is plotted in Fig. \ref{fig: limited FER v. delta}.
Threshold spacing $\Delta = 4$ results in the best FER value. 
Also, the corresponding ANV values are plotted in Fig. \ref{fig: limited ANV v. delta}. 
Above $\Delta = 4$, the threshold spacing values have almost the same ANV results. 
Based on both figures, $\Delta = 4$ is a reasonable choice for threshold spacing value to obtain a fair trade-off between FER performance and ANV. 
 
\begin{figure}[htbp]
\centering
	\includegraphics [scale = .7]{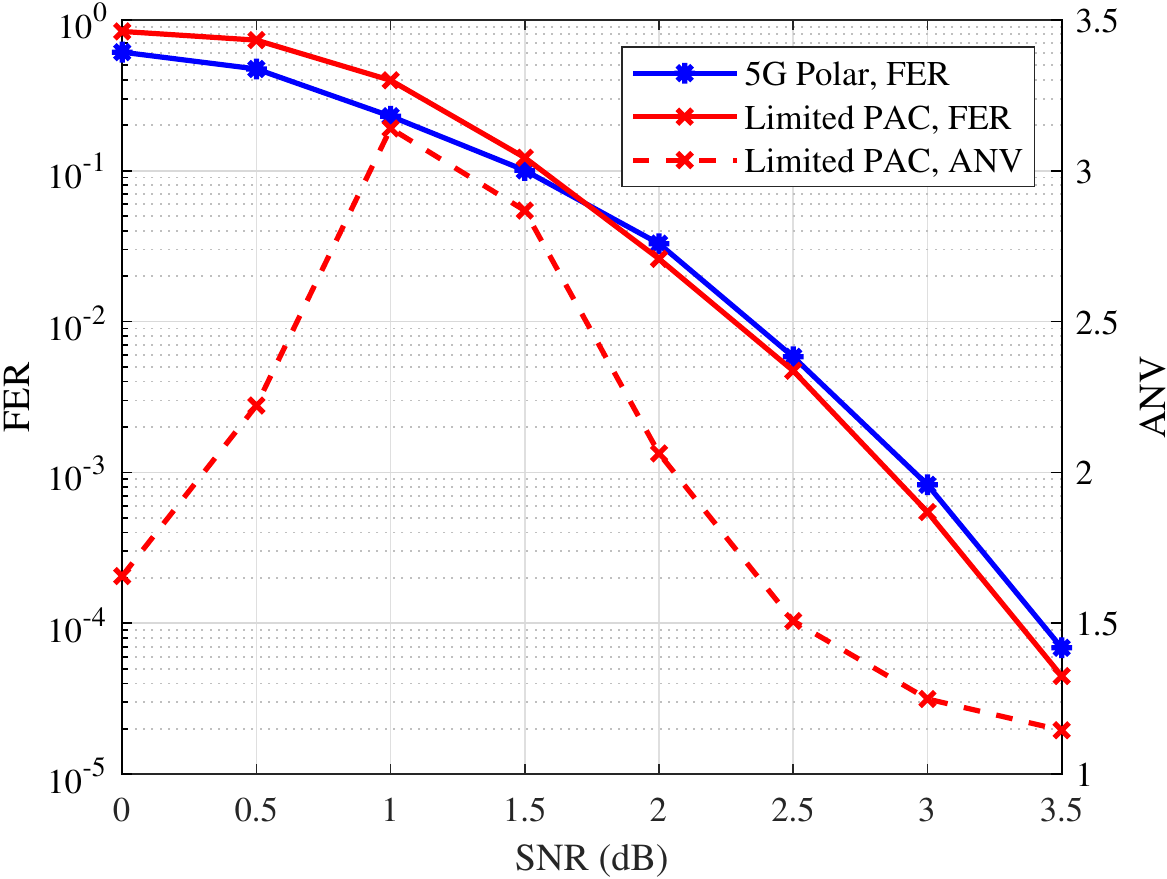}
	\caption{Search-limited PAC code with $\alpha = 0.72$, $\Delta = 4$, and MNV = $2^{12}$ v. polar code with SCL decoding with list size 64 and CRC length of 11.} 
	\label{fig: limited PAC v. 5G}
\end{figure}

Finally, based on the sequential decoding parameters obtained above, in Fig. \ref{fig: limited PAC v. 5G} we compare the error performance of the search-limited PAC code with MNV $ = 2^{12}, \alpha = 0.72 $, and $\Delta = 4$ and SCL decoding of polar codes \cite{tal2011list} with list size of 64 and CRC length of 11. 
It is worth mentioning that for the PAC code, the ANV values are less than 4 for all the SNR values, which is much less than the list size of 64.

%%%%%%%%%%%%%%%%%%%%%%%%%%%%%%%%%%%%%%%%%%%%%%%%%%%%%%%%%%%%%%%%%%%%%%%%%%%%%%%%%%%%%%%%%%%%%%%%%%%%%%%%%%%%%%%%%%%%%%%%%%%%%%%%%%%%%%

\section{Distribution of Computation}\label{sec: Pareto}

\begin{figure}[htbp]
\centering
	\includegraphics [width = \columnwidth]{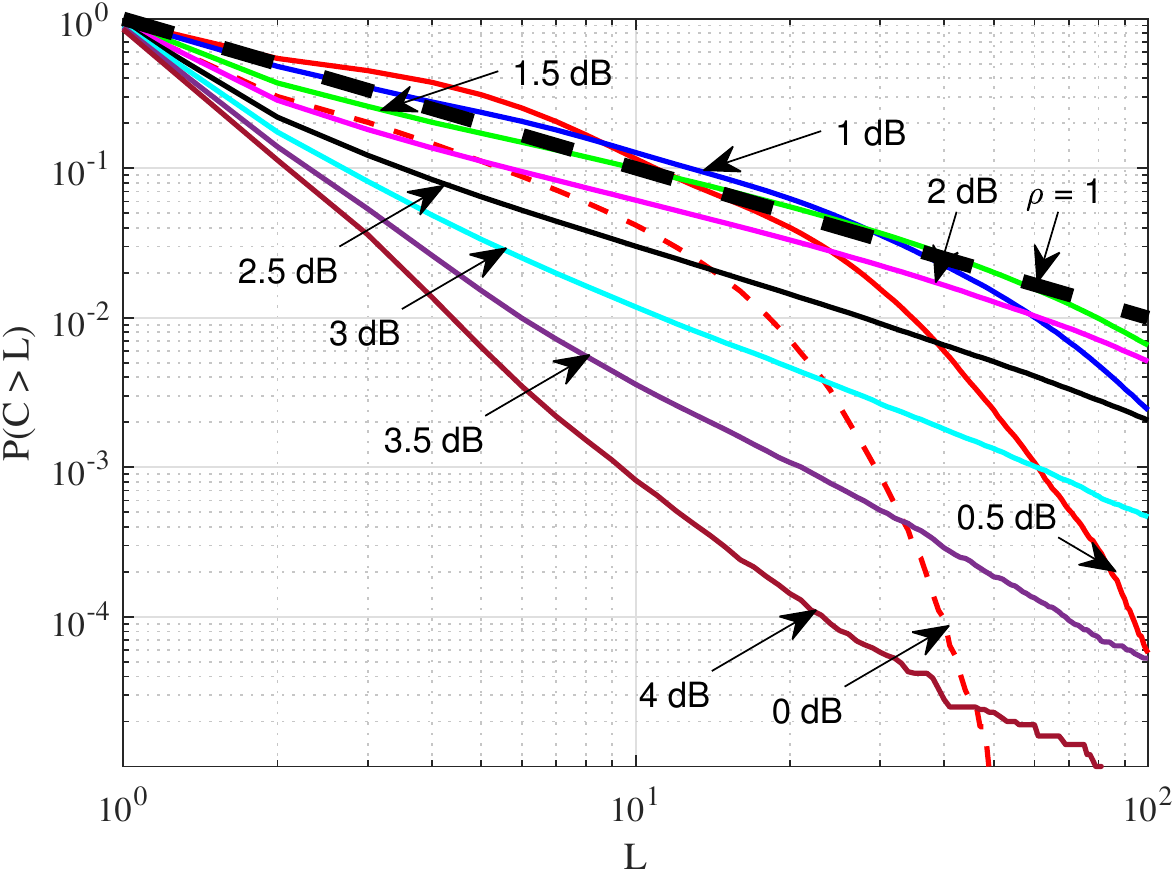}
	\caption{CCDF of the number of node visits of the correctly decoded codewords.} 
	\label{fig: CCDF}
\end{figure}

In this part, we will study the distribution of the number of visits during PAC codes' decoding.
We will only consider the correctly decoded codewords and discard the visit values for the wrongly decoded codewords.
The  complementary cumulative distribution function (CCDF) or survival function of Pareto distribution is defined as
\begin{equation}
    \overline{P}(x) := P(X>L) = \left\{
\begin{array}{ll}
      \sigma_{m}^{\beta}L^{-\beta}, & \text{if}~ L \geq \sigma_m, \\
      1, & \text{if}~ L < \sigma_m,\\
\end{array} 
\right. 
\end{equation}
where the scale parameter $\sigma_m > 0$ is the minimum possible value of $L$, and $\beta > 0$ is the shape parameter. 
Smaller $\beta$ corresponds to more values occurring at the tail of the distribution. 
Pareto distribution at first employed for the distribution of wealth and became known for the 80/20 rule.
This rule says that, as an example, $80\%$ of the wealth is for $20\%$ of the population. 
For the decoding computation of the PAC codes, this rule implies that the decoding computation is typically low but can be high for a small portion of codewords.

If the scale parameter is equal to $1$, the mean of the Pareto distribution is equal to
\begin{equation}
    \mathbb{E}[X] = \left\{
\begin{array}{ll}
      \infty, & \text{if}~ \beta \leq 1, \\
      \frac{\beta}{\beta - 1}, & \text{if}~ \beta > 1.\\
\end{array} 
\right. 
\end{equation}
Pareto distribution has a finite mean for $\beta > 1$ and has a finite variance for $\beta > 2$.
Let $C$ represents the total number of nodes that are visited during a single decoding session.
The CCDF $P(C>L)$ is the probability that the number of visits of a correctly decoded codeword is more than a constant $L$. 
For the sequential decoding of CCs, it is shown in \cite{jacobs} that the distribution of computation required to advance any level is upper bounded by 
\begin{equation}
    P(C_i > L ) < AL^{-\beta},   
\end{equation}
where $A$ and $\beta>0$ are constants, and the exact definition of $C_i$ is given in the next section. 
This proves that the CCDF of the computation time of sequential decoding has a Pareto distribution upper bound. 
For PAC decoding with bias values $b_i = E_0(1,W_N^{(i)})$, the CCDF of the number of nodes visited for the correctly decoded codewords are plotted in Fig. \ref{fig: CCDF} for 1 million decoding trials for different SNR values. 
The results demonstrate that $P(C>L) < L^{-1}$ for almost all SNR values, and consequently, the mean of the upper bound distribution is finite. 
Notice that for the sequential decoding of the CCs, the mean is finite only for the rates below the cutoff rate.

For the SNR values equal to 3.5 and 4 dB, we have $\beta > 2$, which will result in a finite variance.
As an example, Fig. \ref{fig: CCDF} shows that there is one percent probability for a correctly decoded codeword to require more than ten visits per branch at the SNR value of 3.0 dB (outage probability).
The main conclusion from this discussion is that the probability that the number of visits for each codeword exceeds $L$ is going to zero as $L$ increases.

Moreover, for $0$ and $0.5$ dB SNR values, Fig. \ref{fig: CCDF} shows waterfalling results, which means that the outage probability for the low SNR values for a moderate number of computations is small. 
At low SNR values, the PAC decoder finishes the decoding process earlier compared to high SNR values. 
In contrast, for the fixed bias values, PAC codes at low SNR values require exponentially high computation.

\begin{figure}[htbp] 
\centering
	\includegraphics [width = \columnwidth]{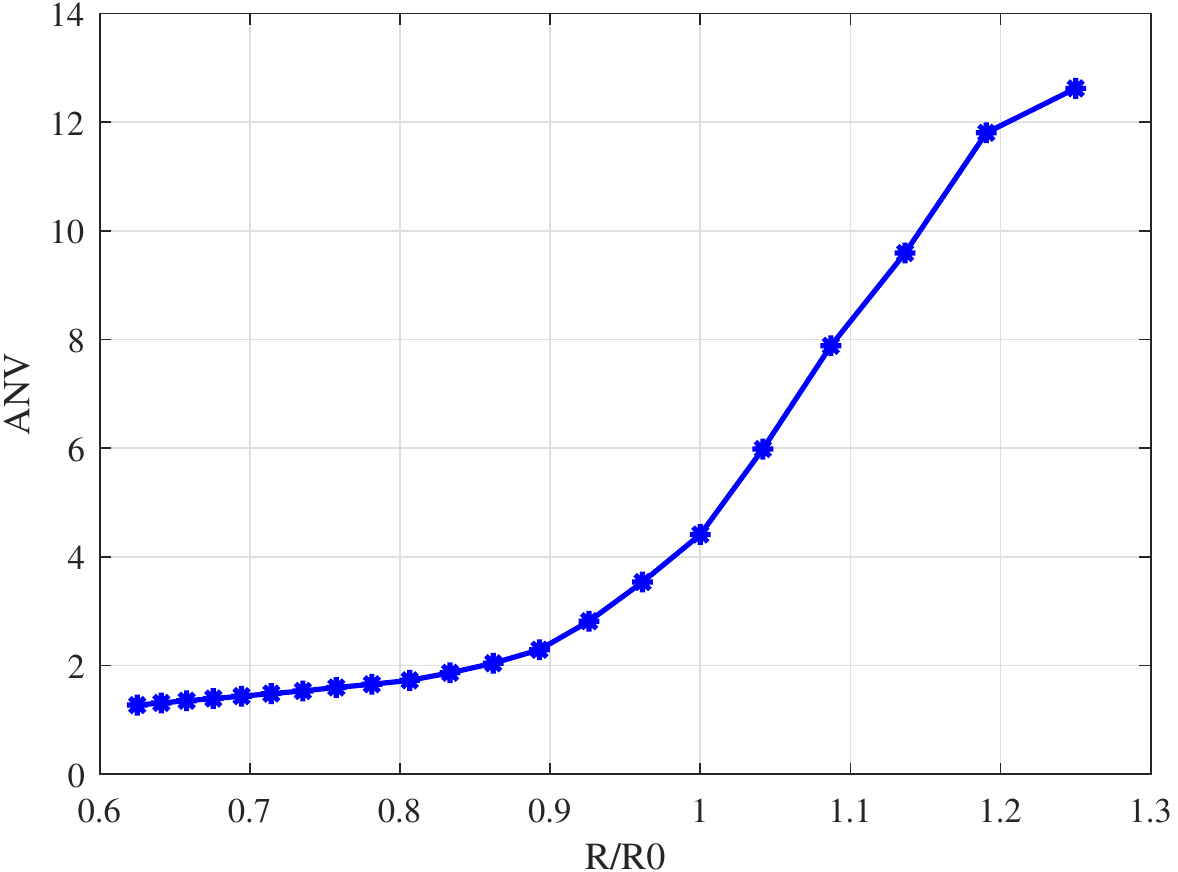}
	\caption{The empirical average number of computations per bit decoded.} 
	\label{fig: Z v. RonR0}
\end{figure}

In Fig. \ref{fig: Z v. RonR0}, for the correctly decoded codewords, the ANV versus $R/R_0$ is plotted, where $R_0$ is the channel cutoff rate. 
As shown in this figure, by increasing $R/R_0$, the ANV increases as well.
Note that around $R/R_0 = 1$, this curve has a mild slope, whereas for the conventional CCs the ANV curve has a sharp slope around $R/R_0 = 1$ \cite{wozencraft1957sequential}.

%%%%%%%%%%%%%%%%%%%%%%%%%%%%%%%%%%%%%%%%%%%%%%%%%%%%%%%%%%%%%%%%%%%%%%%%%%%%%%%%%%%%%%%%%%%%%%%%%%%%%%%%%%%%%%%%%%%%%%%%%%%%%%%%%%%%%%

\section{Upper Bound on Distribution of Computation}\label{sec: upper bound Pareto}
We consider genie-aided sequential decoding of PAC codes. 
Let $\gamma_i$ and $\Tilde{\gamma}_i$ be the $i$th bit metric for the correct and wrong branches, respectively.
In computing $\gamma_i$ and $\Tilde{\gamma}_i$, we assume that the genie provides the correct bit-channel output $(\mathbf{y}, \mathbf{u}^{i-1})$.
Accordingly, the branch metrics for correct and wrong paths can be calculated as
\begin{equation}
  \gamma_i = \log_2 \frac{P(\mathbf{y} , \mathbf{u}^{i-1} | u_i)}{P(\mathbf{y} , \mathbf{u}^{i-1})} - b_i 
\end{equation}
and
\begin{equation}
  \Tilde{\gamma}_i = \log_2 \frac{P(\mathbf{y} , \mathbf{u}^{i-1} | \Tilde{u}_i)}{P(\mathbf{y} , \mathbf{u}^{i-1})} - b_i,
\end{equation}
respectively.
By assuming a genie-aided sequential decoder, decoding the current bit will be independent of what the decoder has decided before. 
After decoding a bit (wrongly or correctly), the decoder will use the correct values of the preceding and current bits provided by the genie to decode the next bit. 
In consequence, the decision of every bit is independent of previously decoded bits. 
If a codeword is not decoded correctly with a genie-aided sequential decoder, it would be decoded wrongly without the help of the genie as well.
The genie reveals the actual value of a bit only after finishing the decoding process of that bit.
For this reason, the actual decoder will make the same decoding errors as the genie-aided decoder.
Hence, $\gamma_i \!\perp\!\!\!\perp \gamma_j$ and $\Tilde{\gamma}_i \!\perp\!\!\!\perp \Tilde{\gamma}_j$ for any $i$ and $j$, and $i \neq j$, where $\!\perp\!\!\!\perp$ is used to represent independency. 

\begin{figure}[htbp]
    \centering
    \begin{tabular}{cc}
    \adjustbox{valign=c}{\subfloat[\label{fig: Incorrect subsets a}]{%
          \includegraphics[width=.45\linewidth]{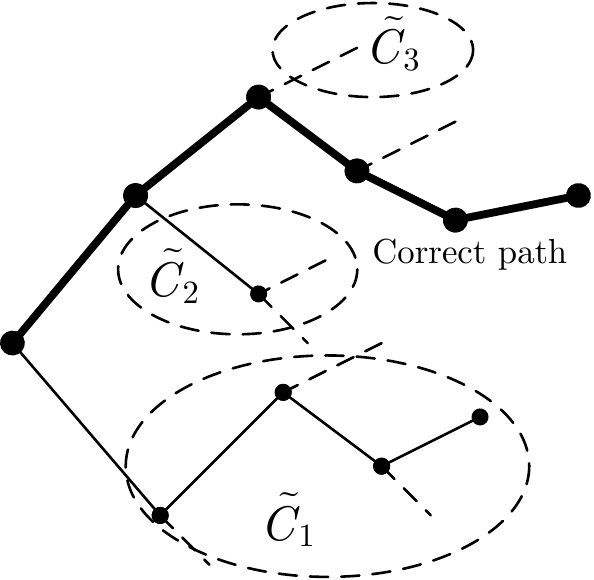}}}
    &      
    \adjustbox{valign=c}{\begin{tabular}{@{}c@{}}
    \subfloat[\label{fig: Incorrect subsets b}]{%
          \includegraphics[width=.45\linewidth]{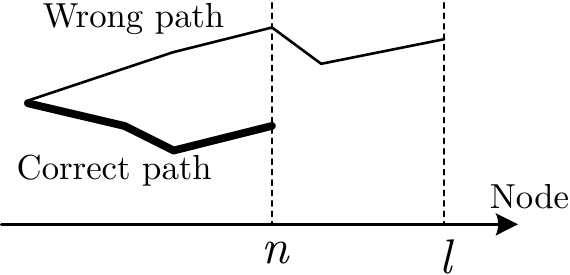}} \\
    \subfloat[\label{fig: Incorrect subsets c}]{%
          \includegraphics[width=.45\linewidth]{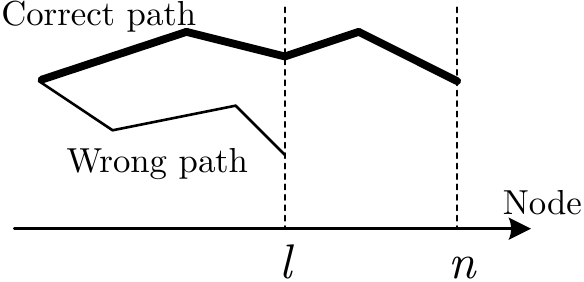}}
    \end{tabular}}
    \end{tabular}
    \caption{Correct path, wrong path, and incorrect subsets on code tree.}
  \label{fig: Incorrect subsets} 
\end{figure}

As Fig. \ref{fig: Incorrect subsets a} displays, we define $\Tilde{C}_i$ as the set of extended nodes in the $i$th incorrect subtree and $C_i$ as the number of computations needed to decode the $i$th correct node. 
We have
\begin{equation}
    C_i = 1 + |\Tilde{C}_i|.
\end{equation}

%%%%%%%%%%%%%%%%%%%%%%%%%%%%%%%%%%%%%%%%%%%%%%%%%%%%%%%%%%%%%%%%%%%%%%%%%%%%%%%%%%%%%%%%%%%%%%%%%%%%%%%%%%%%%%%%%%%%
We use the notation $\Gamma_\text{min}$ as the minimum partial path metric of the correct path and $T_\text{min}$ as the minimum threshold $T$. 
For the Fano algorithm \cite[p.~466]{wozencraft1957sequential}, 
\begin{equation} \label{Gamma_min T_min}
    \Gamma_\text{min} < T_\text{min} + \Delta.
\end{equation}
We label nodes in an incorrect subset of Fig. \ref{fig: Incorrect subsets a} by an ordered pair $(l,m)$ and use $\Tilde{\Gamma}_{l,m}$ for its corresponding partial path metric, where $l$ is the depth of the node and $m$ is its vertical position with any arbitrary order.
When we refer to an arbitrary node in an incorrect set at depth $l$, we use $\Tilde{\Gamma}_{l}$ for its partial path metric. 
We use the notation $\Tilde{\Gamma}_{l,m,\theta}$ for the corresponding metric of node $(l,m)$ in an incorrect set when it is visited for the $\theta$th time. $\Tilde{\Gamma}_{l,m}$ is the metric value when visit is for the first time. $\Tilde{\Gamma}_{l}$ is used as the partial path metric of the $l$th depth for a given incorrect path.

A node $(l,m)$ is able to be extended by the Fano algorithm if its metric satisfies the threshold ($\Tilde{\Gamma}_{l,m} \geq T$).
For any specific threshold $T$, each node can be visited at most once. 
In any revisiting, the threshold is always lower than the previous visit by a $\Delta$.
In summary, for a given $\Tilde{\Gamma}_{l,m}$, the number of visits $\theta$ of a node $(m,l)$ has upper bound \begin{equation}\label{number of visits}
    \theta < \ceil[\big]{\frac{\Tilde{\Gamma}_{l,m} - T_{min}}{\Delta}} \leq \frac{\Tilde{\Gamma}_{l,m} - T_{min}}{\Delta} + 1.
\end{equation}
From (\ref{Gamma_min T_min}) and (\ref{number of visits}) we can conclude that a node $(l,m)$ can be visited for the $\theta$th time if
\begin{equation} \label{ith time visit}
    \Tilde{\Gamma}_{l,m} > \Gamma_{min} + (\theta-2)\Delta.
\end{equation}
Consider a random variable $C_{l,m,\theta}$ with Bernoulli distribution which takes value 1 if node $(m,l)$ is visited for the $\theta$th time ((\ref{ith time visit}) is satisfied). The number of visits $C_1$ has upper bound
\begin{equation} \label{C_1 upper bound}
    C_1 \leq \sum_{l=1}^{\infty}\sum_{m}\sum_{\theta=1}^{\infty}C_{l,m,\theta},
\end{equation}
where $l$ is the depth number, $m$ is the vertical position, and $\theta$ is the number of revisits of the node $(m,l)$. 
In the rest of the paper, we try to obtain an upper bound on the probability of $C_1$ being more than a constant value of $L$.

%%%%%%%%%%%%%%%%%%%%%%%%%%%%%%%%%%%%%%%%%%%%%%%%%%%%%%%%%%%%%%%%%%%%%%%%%%%%%%%%%%%%%%%%%%%%%%%%%%%%%%%%%%%%%%%%%%%%

For a random variable $X$, the moment generating function (MGF) of $X$ is defined as $g(r) := \mathbb{E}[2^{rX}]$ and the semi-invariant MGF of X is defined as $h(r) := \log_2 \big\{\mathbb{E}[2^{rX}] \big\}$.
The following three lemmas will provide upper bounds on the semi-invariant MGF for the bit-channel metrics on the correct and wrong paths and their differences, respectively. 
The proofs are provided in the Appendix. 

\begin{lemma} \label{lemma1}
    Let $h(r_0)$ be the semi-invariant MGF of the bit-channel metric on the correct path. Then 
    \begin{equation}
        h(r_0) := \log_2 \big\{\mathbb{E}[2^{r\gamma_i}] \big\}  \leq -r_0 b_i -(1+r_0)E_0(\frac{-r_0}{1+r_0},W_N^{(i)}),
    \end{equation}
    where $r_0$ is in $(-1,0)$ interval.
\end{lemma}

\begin{lemma} \label{lemma2}
    Let $\Tilde{h}(r)$ be the semi-invariant MGF of the bit-channel metric on the wrong path. We can obtain its upper bound by
    \begin{equation}
        \Tilde{h}(r) := \log_2 \big\{\mathbb{E}[2^{r\Tilde{\gamma}_i}] \big\}  \leq -r b_i -rE_0(\frac{1-r}{r},W_N^{(i)}),
    \end{equation}
    where $r$ is in $(0,1)$ interval.
\end{lemma}

\begin{lemma} \label{lemma3}
    The semi-invariant MGF of the difference of bit-channel metrics on the wrong and the correct paths has an upper bound as
    \begin{equation}
        \log_2 \big\{\mathbb{E}[2^{r(\Tilde{\gamma}_i - \gamma_i)}] \big\}
         \leq -r b_i -rE_0(\frac{1-r}{r},W_N^{(i)}),
    \end{equation}
    where $r$ is in $(0,1)$ interval, and it is assumed that the bias $b_i \leq \frac{E_0(\delta,W_N^{(i)})}{\delta} $ for $0<\delta<1$.
\end{lemma}

With the help of Wald's identity, we prove the following lemma.
The lemma says that the probability of $\Gamma_\text{min}$ being less than a constant value goes exponentially fast to zero. 
A small value of $\Gamma_\text{min}$ will result in the chance of the partial path metric on the wrong path being greater than $\Gamma_\text{min}$, which means that the decoder will advance further in the wrong directions. 

\begin{lemma}
    The probability that the minimum partial path metric on the correct path is less than a constant absorbing barrier $\mu$ is upper bounded as
    \begin{equation}
        P(\Gamma_\text{min} \leq \mu ) \leq 2^{-r_0 \mu},
    \end{equation}
    where $r_0$ is in $(-1,0)$ interval and it is assumed that the bias $b_i \leq \frac{E_0(\delta,W_N^{(i)})}{\delta} $ for $0<\delta<1$.
\end{lemma}

Suppose that the correct path segment is shorter than the wrong path segment (i.e. $n < l$) as shown in Fig. \ref{fig: Incorrect subsets b}. The following lemma is useful to provide an upper bound on the probability that the wrong path will be extended further when $n < l$.
\begin{lemma} \label{lemma5}
Assuming $n < l$, we have
\begin{equation}
\begin{split}
    &P(\Tilde{\Gamma}_l \geq \Gamma_n + \alpha) 
     \leq 2^{-r \alpha} 2^{-r \sum_{i =1}^{l}\left[ E_0(\frac{1-r}{r},W_N^{(i)}) + b_i \right]},
\end{split}
\end{equation}
where $0<r<1$ and $b_i \leq \frac{E_0(\delta,W_N^{(i)})}{\delta} $ for $0<\delta<1$.
As an special case, assume that $\alpha = 0$, $r = \frac{1}{2}$, and $b_i = E_0(1,W_N^{(i)})$.
Then, $P(\Tilde{\Gamma}_l \geq \Gamma_n) $ goes exponentially to zero with an exponent equal to $\sum_{i = 1}^{l}E_0(1,W_N^{(i)})$.
\end{lemma}

Next assume that the correct path segment advances more than the wrong path segment (i.e. $n \geq l$) as shown in Fig. \ref{fig: Incorrect subsets c}. The following lemma will provide an upper bound on the probability that the wrong path will be extended further when $n \geq l$.
\begin{lemma} \label{lemma6}
By assuming that $n \geq l$, we have
\begin{equation}
\begin{split}
    &P(\Tilde{\Gamma}_l \geq \min_{n\geq l}\{\Gamma_n\} + \alpha)
     \leq 2^{r_0 \alpha} 2^{r_0 \sum_{i =1}^{l}( E_0(\frac{1+r_0}{-r_0},W_N^{(i)}) + b_i)},
\end{split}
\end{equation}
where $r_0 \in (-1,0)$ and $b_i \leq \frac{E_0(\delta,W_N^{(i)})}{\delta} $ for $0<\delta<1$.
Similar to Lemma \ref{lemma5}, as an special case, assume that $\alpha = 0$, $r_0 = \frac{-1}{2}$, and $b_i = E_0(1,W_N^{(i)})$.
Then, $P(\Tilde{\Gamma}_l \geq \min\{\Gamma_n\}) $ goes exponentially to zero with an exponent equal to $\sum_{i = 1}^{l}E_0(1,W_N^{(i)})$.
\end{lemma}
Whenever the partial path metric of any incorrect path at a given depth is above the $\Gamma_\text{min}$, the wrong direction finds the chance to continue more. 
The following theorem finds an upper bound on the probability that latter can happen.
\begin{theorem}\label{theorem 1}
The probability that the partial path metric of the wrong path is greater than or equal to the minimum partial path of the correct path by a constant $\alpha$ is upper bounded as
\begin{equation}
\begin{split}
    & Pr\left[ \Tilde{\Gamma}_l \geq \Gamma_\text{min} + \alpha \right] 
    \leq (l+1) 2^{-r\alpha} 2^{-r \sum_{i =1}^{l}\left[ E_0(\frac{1-r}{r},W_N^{(i)}) + b_i \right]},
\end{split}
\end{equation}
where it is assumed that $r \in (0,1)$ and $b_i < \frac{E_0(\delta,W_N^{(i)})}{\delta} $ for $0<\delta<1$.
Similarly for $\alpha = 0$, $r = \frac{1}{2}$, and $b_i = E_0(1,W_N^{(i)})$, the upper bound exponent is $\sum_{i = 1}^{l}E_0(1,W_N^{(i)})$, and the chance of advancing more than $l$ steps in the wrong path has upper bound which is a linear function of $l$.
\end{theorem}

Suppose that the decoder is in the $(l,m)$th node of the decoding tree, where out of the first $i$ depths $\lambda_i$ of them are information bits (the nodes for which the tree branches) s.t. $1\leq i\leq l$. 
We define partial rate as $R_i = \frac{\lambda_i}{i}$.
We can easily see that $2^{\lambda_l} = 2^{lR_l}$ is an upper bound on the number of incorrect nodes at depth $l$ of the tree. 

Furthermore, assume that 
\begin{equation} \label{partial rate}
    \sum_{i=1}^{l}R_i \leq r\sum_{i=1}^{l}\left(  E_0(\frac{1-r}{r},W_N^{(i)})+b_i \right) - \epsilon,
\end{equation}
where $\epsilon$ is a small positive number. 
We would like to mention that by substituting the parameters used in our simulations, $r = \frac{1}{2}$ and $b_i = E_0(1,W_N^{(i)})$, the inequality reduces to
\begin{equation}
    \sum_{i=1}^{l} R_i \leq \sum_{i=1}^{l} E_0(1,W_N^{(i)}) -\epsilon.
\end{equation}
%the sum of the first $l$ bit-channel cutoff rates are less than the number of the information bits in the first $l$ bits as

With the conditions mentioned above, the following theorem will give an upper bound for the $\mathbb{E}[C_1],$ which corresponds to the average number of computations needed to decode the first bit. The average is over data sequence, the channel noise, and the ensemble of the PAC codes.

By (\ref{C_1 upper bound}), we have
\begin{equation} \label{expected C0}
\begin{split}
    & \mathbb{E}[C_1] \leq \sum_{l=1}^{\infty}\sum_{m}\sum_{\theta=1}^{\infty} \mathbb{E}[C_{l,m,\theta}] \\
    & = \sum_{l=1}^{\infty}\sum_{m}\sum_{\theta=1}^{\infty}
    P(\Tilde{\Gamma}_{l,m} > \Gamma_{min} + (\theta - 2)\Delta),
\end{split}
\end{equation}
where $C_{l,m,\theta}$ has a Bernoulli distribution with probability of being one equal to $P(\Tilde{\Gamma}_{l,m} > \Gamma_{min} + (\theta - 2)\Delta)$.

\begin{theorem}
The value $\Delta = \frac{1}{r}$ minimizes the upper bound of $\mathbb{E}[C_1]$ to
\begin{equation}
    \mathbb{E}[C_1] \leq  
    \frac{4}{(1-2^{-\epsilon})^2},
\end{equation}
where $\epsilon$ satisfies (\ref{partial rate}).
\end{theorem}

Ultimately, the following theorem gives an upper bound on the CCDF of the bit-channel computations.

\begin{theorem}
Suppose that $C_n$ is the number of computations required to decode the $n$th bit. The probability that $C_n$ is greater than a constant value $L$ has a Pareto distribution upper bound as
\begin{equation}
\begin{split}
    & P(C_n \geq L) \leq \frac{\mathbb{E}[C_n^{\beta}]}{L^{\beta}}
    \leq 
     \left(\frac{4}{ L(1-2^{\frac{-\epsilon}{\beta}})^2}\right)^{\beta},
\end{split}
\end{equation}
where $\beta > 1$. To proof this we assume that $b_i \leq \frac{E_0(\delta,W_N^{(i)})}{\delta}$ for $0<\delta<1$ and 
\begin{equation} \label{partial rate}
    \sum_{i=1}^{l}R_i \leq r\sum_{i=1}^{l}\left(  E_0(\frac{1-r}{r},W_N^{(i)})+b_i \right) - \epsilon,
\end{equation}
where $0<r<1$ and $\epsilon > 0$.
\end{theorem}

%%%%%%%%%%%%%%%%%%%%%%%%%%%%%%%%%%%%%%%%%%%%%%%%%%%%%%%%%%%%%%%%%%%%%%%%%%%%%%%%%%%%%%%%%%%%%%%%%%%%%%%

\section{Conclusion}\label{sec: conclusion}

In this paper, we derived an optimal metric function on average that uses the bit-channel mutual information as the bias value, which results in a favorable trade-off between the performance and the computation of the sequential decoding of PAC codes. 
Moreover, we introduced a construction to find good values of the threshold spacing and scaling parameter of the bias to improve the error-correction and the computation trade-off for the search-limited and -unlimited PAC codes. 
Finally, we proved that by using the bias values less the bit-channel cutoff rates, the PAC codes' sequential decoding has a Pareto distribution upper bound on its computations; this proves the probability of having a high computation in decoding goes to zero.
Using the bit-channel capacity or cutoff rate as the bit-channel bias values, simulation results demonstrated that the PAC codes' superior error-correction performance was kept while benefiting from the sequential decoding's low computation.

Finding a tight error-performance upper bound for the PAC codes is a future study.

%%%%%%%%%%%%%%%%%%%%%%%%%%%%%%%%%%%%%%%%%%%%%%%%%%%%%%%%%%%%%%%%%%%%%%%%%%%%%%%%%%%%%%%%%%%%%%%%%%%%%%%
% \begin{equation*}
%  dgfb \IEEEQEDhereeqn\quad   
% \end{equation*}
% dfg \hfill\IEEEQEDhere 

\appendix \label{Appendix}

\subsection{Proof of Lemma 1}
\begin{equation}
\begin{split}
    & g(r_0) := \mathbb{E}[2^{r_0\gamma_i}] = 
    \mathbb{E}\left[2^{r_0\frac{P(\mathbf{y} , \mathbf{u}^{i-1} | u_i)}{P(\mathbf{y} , \mathbf{u}^{i-1})} - r_0b_i } \right] \\
    & = \mathbb{E}\left[ \left(\frac{P(\mathbf{y} , \mathbf{u}^{i-1} | u_i)}{P(\mathbf{y} , \mathbf{u}^{i-1})} \right)^{r_0}2^{-r_0b_i} \right]\\
    & = \sum_{u_i}q(u_i)\sum_{(\mathbf{y},\mathbf{u}^{i-1})} P(\mathbf{y} , \mathbf{u}^{i-1} | u_i)
    \left(\frac{P(\mathbf{y} , \mathbf{u}^{i-1} | u_i)}{P(\mathbf{y} , \mathbf{u}^{i-1})} \right)^{r_0}2^{-r_0b_i}\\
    & = 2^{-r_0b_i}\sum_{(\mathbf{y},\mathbf{u}^{i-1})} \underbrace{P(\mathbf{y},\mathbf{u}^{i})^{-r_0}}_a
    \underbrace{\sum_{u_i} q(u_i) P(\mathbf{y} , \mathbf{u}^{i-1} | u_i)^{1+r_0}}_b.
\end{split}
\end{equation}

By defining $s = -r_0$ and considering $-1 < r_0 < 0$ and using the Cauchy-Schwarz (CS) inequality
\begin{equation}
    \sum ab \leq \left(\sum a^{\frac{1}{s}} \right)^{s}
    \left(\sum b^{\frac{1}{1-s}} \right)^{1-s},
\end{equation}
we have
\begin{equation}
\begin{split}
    & g(r_0) 
    \overset{\text{CS}}{\le}
    2^{-r_0 b_i} \left[\underbrace{\sum_{(\mathbf{y},\mathbf{u}^{i-1})}P(\mathbf{y},\mathbf{u}^{i})}_\text{= 1} \right]^{s}\\
    &~~~~~~~~~ \left[\sum_{(\mathbf{y},\mathbf{u}^{i-1})} \left[\sum_{u_i} q(u_i) P(\mathbf{y} , \mathbf{u}^{i-1} | u_i)^{1-s} \right]^{\frac{1}{1-s}} \right]^{1-s}\\
    & = 2^{-r_0 b_i} 2^{(1-s)\log_2 \left[\sum_{(\mathbf{y},\mathbf{u}^{i-1})} \left[\sum_{u_i} q(u_i) P(\mathbf{y} , \mathbf{u}^{i-1} | u_i)^{1-s} \right]^{\frac{1}{1-s}} \right]}\\
    & = 2^{-r_0b_i -(1-s)E_0(\frac{s}{1-s},W_N^{(i)})}
     = 2^{-r_0b_i -(1+r_0)E_0(\frac{-r_0}{1+r_0},W_N^{(i)})}.\\
\end{split}
\end{equation}
By taking $\log_2$ function, we conclude
\begin{equation}
    h(r_0) \leq -r_0b_i -(1+r_0)E_0(\frac{-r_0}{1+r_0},W_N^{(i)}).
\end{equation}
\hfill\IEEEQEDhere

%%%%%%%%%%%%%%%%%%%%%%%%%%%%%%%%%%%%%%%%%%%%%%%%%%%%%%%%%%%%%%%%%%%%%%%%%%%%%%%%%%%%%%%%%%%%%%%%%%%%%%%
\subsection{Proof of Lemma 2}
Consider the MGF $g(\Tilde{\gamma}_i)$.
\begin{equation}
\begin{split}
 & g(\Tilde{\gamma}_i) = \mathbb{E}[2^{r\Tilde{\gamma}_i}] = \mathbb{E}\left[2^{r\frac{P(\mathbf{y} , \mathbf{u}^{i-1} | \Tilde{u}_i)}{P(\mathbf{y} , \mathbf{u}^{i-1})} - rb_i } \right] \\
    & = \mathbb{E}\left[ \left(\frac{P(\mathbf{y} , \mathbf{u}^{i-1} | \Tilde{u}_i)}{P(\mathbf{y} , \mathbf{u}^{i-1})} \right)^{r}2^{-rb_i} \right]\\
    & = \sum_{\Tilde{u}_i}\sum_{(\mathbf{y},\mathbf{u}^{i-1})}
    \underbrace{\sum_{u_i}q(u_i) P(\mathbf{y} , \mathbf{u}^{i-1} | u_i)}
    _\text{=  $P(\mathbf{y} , \mathbf{u}^{i-1})$} q(\Tilde{u}_i)\\
    &~~~~~ \left(\frac{P(\mathbf{y} , \mathbf{u}^{i-1} | \Tilde{u}_i)}
    {P(\mathbf{y} , \mathbf{u}^{i-1})} \right)^{r}2^{-rb_i}\\
    & = 2^{-rb_i} \sum_{(\mathbf{y},\mathbf{u}^{i-1})} 
    \underbrace{P(\mathbf{y} , \mathbf{u}^{i-1})^{1-r}}_b
    \underbrace{\sum_{\Tilde{u}_i}q(\Tilde{u}_i)P(\mathbf{y} , \mathbf{u}^{i-1} | \Tilde{u}_i)^{r}}_a.
\end{split}
\end{equation}
For $0<r<1$, and using the Cauchy-Schwarz inequality as
\begin{equation}
    \sum ab \leq \left(\sum a^{\frac{1}{r}} \right)^{r}
    \left(\sum b^{\frac{1}{1-r}} \right)^{1-r},
\end{equation}
we obtain
\begin{equation}
\begin{split}
    & g(\Tilde{\gamma}_i) \leq 2^{-rb_i}\left[\underbrace{\sum_{(\mathbf{y},\mathbf{u}^{i-1})} P(\mathbf{y} , \mathbf{u}^{i-1})}_\text{= 1} \right]^{1-r} \\
    &~~~~~~~~~ \left[\sum_{(\mathbf{y},\mathbf{u}^{i-1})} \left[\sum_{\Tilde{u}_i} q(\Tilde{u}_i) P(\mathbf{y} , \mathbf{u}^{i-1} | \Tilde{u}_i)^{r} \right]^{\frac{1}{r}} \right]^{r}\\
    & = 2^{-rb_i} 2^{r\log_2\left\{\sum_{(\mathbf{y},\mathbf{u}^{i-1})} \left[\sum_{\Tilde{u}_i} q(\Tilde{u}_i) P(\mathbf{y} , \mathbf{u}^{i-1} | \Tilde{u}_i)^{r} \right]^{\frac{1}{r}} \right\} }\\
    & = 2^{-r b_i -rE_0(\frac{1-r}{r},W_N^{(i)})}.
\end{split}
\end{equation}
Finally, by taking the $\log_2$ function from both sides, the result becomes
\begin{equation}
        \Tilde{h}(r) = \log_2 g(\Tilde{\gamma}_i)  \leq -r b_i -rE_0(\frac{1-r}{r},W_N^{(i)}).
\end{equation}
\hfill\IEEEQEDhere 

%%%%%%%%%%%%%%%%%%%%%%%%%%%%%%%%%%%%%%%%%%%%%%%%%%%%%%%%%%%%%%%%%%%%%%%%%%%%%%%%%%%%%%%%%%%%%%%%%%%%%%%

\subsection{Proof of Lemma 3}
\begin{equation}
\begin{split}
    &\mathbb{E}[2^{r(\Tilde{\gamma}_i - \gamma_i)}] = 
    \sum_{u_i}\sum_{(\mathbf{y},\mathbf{u}^{i-1})}\sum_{\Tilde{u}_i}
    q(u_i)P(\mathbf{y} , \mathbf{u}^{i-1} | u_i)q(\Tilde{u}_i)\\
    &~~~~~~~~~~~~~~~~~~ 2^{r(\Tilde{\gamma}_i - \gamma_i)}\\
    &= \sum_{u_i}\sum_{(\mathbf{y},\mathbf{u}^{i-1})}\sum_{\Tilde{u}_i}
    q(u_i)P(\mathbf{y} , \mathbf{u}^{i-1} | u_i)q(\Tilde{u}_i)\\
    &~~~~ \left[\frac{P(\mathbf{y} , \mathbf{u}^{i-1} | \Tilde{u}_i)}{\cancel{P(\mathbf{y} , \mathbf{u}^{i-1})}} \right]^{r}
    \left[\frac{\cancel{P(\mathbf{y} , \mathbf{u}^{i-1})}}{P(\mathbf{y} , \mathbf{u}^{i-1} | u_i)} \right]^{r}\\
    &= \sum_{u_i}\sum_{(\mathbf{y},\mathbf{u}^{i-1})}\sum_{\Tilde{u}_i}
    q(u_i)P(\mathbf{y} , \mathbf{u}^{i-1} | u_i)^{1-r}q(\Tilde{u}_i)P(\mathbf{y} , \mathbf{u}^{i-1} | \Tilde{u}_i)^{r}\\
    &= \sum_{(\mathbf{y},\mathbf{u}^{i-1})} 
    \underbrace{\sum_{u_i}q(u_i)P(\mathbf{y} , \mathbf{u}^{i-1} | u_i)^{1-r}}_\text{= $a$}  
     \underbrace{\sum_{\Tilde{u}_i}q(\Tilde{u}_i)P(\mathbf{y} , \mathbf{u}^{i-1} | \Tilde{u}_i)^{r}}_\text{= $b$} \\
    & \overset{\text{CS}}{\le}
    \left[\sum_{(\mathbf{y},\mathbf{u}^{i-1})}\left[\sum_{u_i}q(u_i)P(\mathbf{y} , \mathbf{u}^{i-1} | u_i)^{1-r} \right]^{\frac{1}{1-r}} \right]^{1-r}\\
    &~~~~ \left[\sum_{(\mathbf{y},\mathbf{u}^{i-1})}\left[\sum_{\Tilde{u}_i}q(\Tilde{u}_i)P(\mathbf{y} , \mathbf{u}^{i-1} | \Tilde{u}_i)^{r} \right]^{\frac{1}{r}} \right]^{r}\\
    &= 2^{(1-r)\log_2\left[\sum_{(\mathbf{y},\mathbf{u}^{i-1})}\left[\sum_{u_i}q(u_i)P(\mathbf{y} , \mathbf{u}^{i-1} | u_i)^{1-r} \right]^{\frac{1}{1-r}} \right]}\\
    &~~~~ 2^{r\log_2\left[\sum_{(\mathbf{y},\mathbf{u}^{i-1})}\left[\sum_{\Tilde{u}_i}q(\Tilde{u}_i)P(\mathbf{y} , \mathbf{u}^{i-1} | \Tilde{u}_i)^{r} \right]^{\frac{1}{r}} \right]}\\
    & = 2^{-(1-r)E_0(\frac{r}{1-r},W_N^{(i)})}
    2^{-rE_0(\frac{1-r}{r},W_N^{(i)})}.
\end{split}
\end{equation}
Assume that the bias $b_i < \frac{1-r}{r}E_0(\frac{r}{1-r},W_N^{(i)})$ for $0<r<1$ or equivalently $b_i < \frac{E_0(\delta,W_N^{(i)})}{\delta}$ for $0< \delta = \frac{r}{1-r}<1$.
As a result, we have
\begin{equation}
    \mathbb{E}[2^{r(\Tilde{\gamma}_i - \gamma_i)}] \leq
    2^{-rb_i -rE_0(\frac{1-r}{r},W_N^{(i)})}.
\end{equation}
We can conclude the lemma by taking monotonically increasing $\log_2$ function from both sides of the inequality.

\hfill\IEEEQEDhere 

%%%%%%%%%%%%%%%%%%%%%%%%%%%%%%%%%%%%%%%%%%%%%%%%%%%%%%%%%%%%%%%%%%%%%%%%%%%%%%%%%%%%%%%%%%%%%%%%%%%%%%%

\subsection{Proof of Lemma 4}
\begin{figure}[htbp] 
\centering
	\includegraphics[width = 0.8\columnwidth]{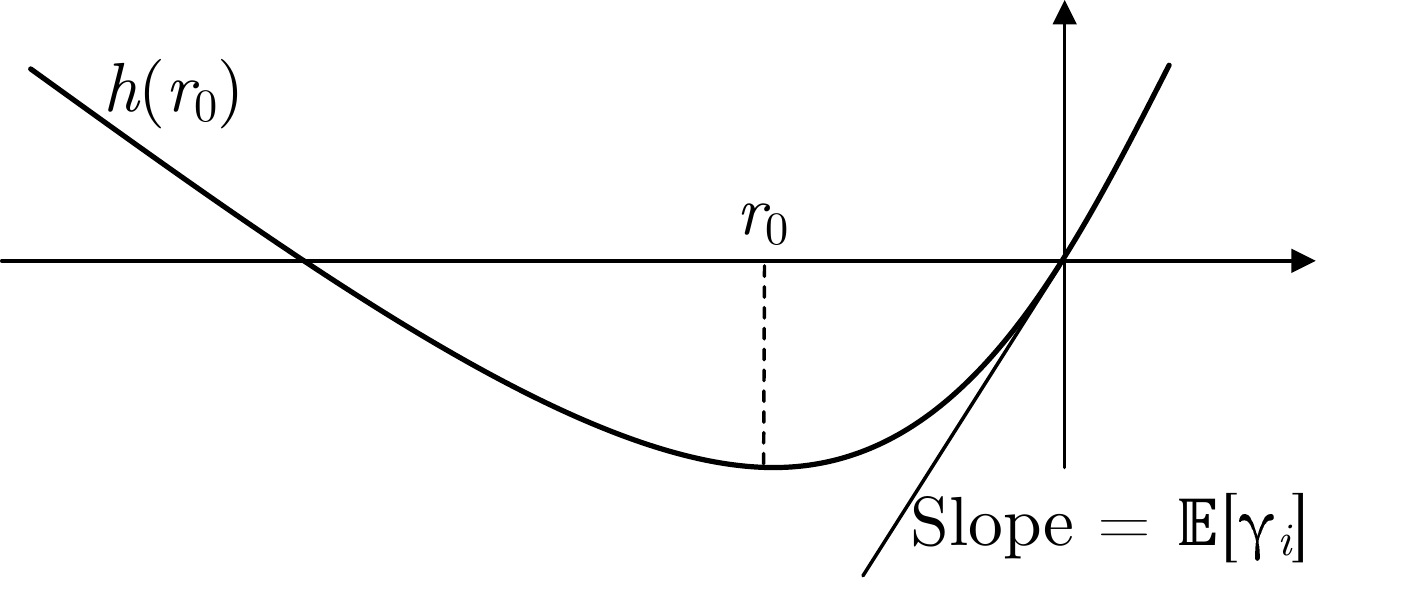}
	\caption{Semi-invariant MGF.} 
	\label{fig: MGF}
\end{figure}

By taking the derivative of the semi-invariant MGF $h(r_0)$ we have
\begin{equation}
   h^{'}(r_0) = \frac{g^{'}(r_0)}{g(r_0)} = \frac{\mathbb{E}[\gamma_i]}{g(r_0)}.
\end{equation}
We know that $g(0) = 1$, and at the origin we have that $h^{'}(0) = \mathbb{E}[\gamma_i]$. By taking the derivative of 
\begin{equation}
\begin{split}
    & g(r_0) = \mathbb{E}[2^{r_0\gamma_i}]  
     \\
    & = \sum_{u_i}q(u_i)\sum_{(\mathbf{y},\mathbf{u}^{i-1})} P(\mathbf{y} , \mathbf{u}^{i-1} | u_i)
    \left(\frac{P(\mathbf{y} , \mathbf{u}^{i-1} | u_i)}{P(\mathbf{y} , \mathbf{u}^{i-1})} \right)^{r_0}2^{-r_0b_i},
\end{split}
\end{equation}
we have
\begin{equation}
\begin{split}
    & g^{'}(r_0) = \sum_{u_i}q(u_i)\sum_{(\mathbf{y},\mathbf{u}^{i-1})} P(\mathbf{y} , \mathbf{u}^{i-1} | u_i)\\
    &~~~~ \log_2\left(\frac{P(\mathbf{y} , \mathbf{u}^{i-1} | u_i)}{P(\mathbf{y} , \mathbf{u}^{i-1})}  \right)
    \left(\frac{P(\mathbf{y} , \mathbf{u}^{i-1} | u_i)}{P(\mathbf{y} , \mathbf{u}^{i-1})} \right)^{r_0}2^{-r_0b_i} \ln(2)\\
    &~~~~ - \sum_{u_i}q(u_i)\sum_{(\mathbf{y},\mathbf{u}^{i-1})} P(\mathbf{y} , \mathbf{u}^{i-1} | u_i)\\
    &~~~~ \left(\frac{P(\mathbf{y} , \mathbf{u}^{i-1} | u_i)}{P(\mathbf{y} , \mathbf{u}^{i-1})} \right)^{r_0}2^{-r_0b_i}b_i \ln(2).\\
\end{split}
\end{equation}
As a result, for $r_0 = 0$ we have
\begin{equation}
\begin{split}
    & g^{'}(0) = \sum_{u_i}q(u_i)\sum_{(\mathbf{y},\mathbf{u}^{i-1})} P(\mathbf{y} , \mathbf{u}^{i-1} | u_i)\\
    &~~~~~~~~~~ \log_2\left(\frac{P(\mathbf{y} , \mathbf{u}^{i-1} | u_i)}{P(\mathbf{y} , \mathbf{u}^{i-1})}  \right) \ln(2)\\
    &~~~~~~~~~~ - \sum_{u_i}q(u_i)\sum_{(\mathbf{y},\mathbf{u}^{i-1})} P(\mathbf{y} , \mathbf{u}^{i-1} | u_i)  b_i \ln(2)\\
    &= I(W_N^{(i)})\ln(2) - b_i \ln(2).
\end{split}
\end{equation}
A random variable has a negative drift when its expectation is negative and has a positive drift when its expectation is positive. 
We see that $h(0) = \log_2(g(0)) = 0$, and
\begin{equation}
    h^{'}(0) = \mathbb{E}[\gamma_{i}] = g^{'}(0) > 0 \ \ \ \ \ \text{iff} \ \ \ \ b_i < I(W_N^{(i)}).
\end{equation}
Using the upper bound derived in Lemma \ref{lemma1}, it can be affirmed that for an $-1<r_0<0$, the semi-invariant MGF $h(r_0)$ becomes negative if and only if
\begin{equation}
    h(r_0) \leq -r_0 b_i -(1+r_0)E_0(\frac{-r_0}{1+r_0},W_N^{(i)}) < 0,
\end{equation}
where this occurs if and only if
\begin{equation}
    b_i < \frac{1+r_0}{-r_0}E_0(\frac{-r_0}{1+r_0},W_N^{(i)}).
\end{equation}
So, we can conclude that $h(r_0) < 0$ if and only if
\begin{equation}
    b_i < \frac{E_0(\delta,W_N^{(i)})}{\delta}~~~~ s.t.~~~
    \delta := \frac{-r_0}{1+r_0}, ~~~ 0< \delta < 1.
\end{equation}
Fig. \ref{fig: MGF} shows a typical behaviour of the semi-invariant MGF $h(r)$. 
Because $P(\gamma_i > 0) > 0$ and $P(\gamma_i < 0) > 0$, obviously we can see that $h(r) \longrightarrow \infty$  from both sides.
We are now equipped to use Wald's identity \cite[p.~434]{gallager2013stochastic} to conclude the proof:\\

\textbf{Wald's identity.} 
Let $\{ \gamma_i; i \geq 1\}$ be IID r.v's and $h(r) = \log_2\{\mathbb{E}[2^{r\gamma_i}]\}$ be the MGF of each $\gamma_i$. Let $\Gamma_j = \sum_{i = 1}^{j} \gamma(u_i; \mathbf{y}, \mathbf{u}^{i-1})$ and $\Gamma_\text{min} = \inf \Gamma_j$. Then for any $r_0 < 0$ s.t. $h(r_0) \leq 0$, and any absorbing barrier $\mu$,
\begin{equation}
    P (\Gamma_{min} < \mu) \leq 2^{- r_0 \mu},
\end{equation}
where $\Gamma_{min} = \inf \Gamma_L$ is the infimum of the partial path metric values on the correct path.

\hfill\IEEEQEDhere 

%%%%%%%%%%%%%%%%%%%%%%%%%%%%%%%%%%%%%%%%%%%%%%%%%%%%%%%%%%%%%%%%%%%%%%%%%%%%%%%%%%%%%%%%%%%%%%%%%%%%%%%

\subsection{Proof of Lemma 5}
It is assumed that $l > n$. 
\begin{equation}
\begin{split}
    &P(\Tilde{\Gamma}_l \geq \Gamma_n + \alpha) 
     = P(2^{r \Tilde{\Gamma}_l} \geq 2^{r(\Gamma_n + \alpha)}) \\
    & \overset{\text{CB}}{\le}
     \mathbb{E} \left[ 2^{r\left[\sum_{i=1}^{l}\Tilde{\gamma}_i - \sum_{i=1}^{n}\gamma_i -\alpha \right]} \right] \\
    &= 2^{-r\alpha} \prod_{i=1}^{n} \mathbb{E}\left[ 2^{r(\Tilde{\gamma}_i - \gamma_i)} \right] \prod_{i=n+1}^{l} \mathbb{E}\left[ 2^{r\Tilde{\gamma}_i } \right] \\
    & \leq 2^{-r \alpha}
    2^{-r \sum_{i =1}^{n}\left[ E_0(\frac{1-r}{r},W_N^{(i)}) + b_i \right]}
    2^{-r \sum_{i =n+1}^{l}\left[ E_0(\frac{1-r}{r},W_N^{(i)}) + b_i \right]}\\
    & = 2^{-r \alpha} 2^{-r \sum_{i =1}^{l}\left[ E_0(\frac{1-r}{r},W_N^{(i)}) + b_i \right]},
\end{split}
\end{equation}
where the first inequality is by Chernoff-bound (CB), the second equality is by the genie aided decoding assumption, and the second inequality is by Lemma \ref{lemma2} and \ref{lemma3}.

\hfill\IEEEQEDhere 

%%%%%%%%%%%%%%%%%%%%%%%%%%%%%%%%%%%%%%%%%%%%%%%%%%%%%%%%%%%%%%%%%%%%%%%%%%%%%%%%%%%%%%%%%%%%%%%%%%%%%%%

\subsection{Proof of Lemma 6}
\noindent It is assumed that $l \leq n$. Furthermore, we define 
\begin{equation}
    \Gamma_{l+1}^n := \sum_{i=l+1}^{n}\gamma_{i}.
\end{equation}
when $n>l$, and $\Gamma_{l+1}^n = 0$ when $n = l$.
Then, in this case ($l \leq n$), we have that
\begin{equation}
    \Gamma_\text{min} = \Gamma_l + \inf_{\forall n \geq l} \{\Gamma_{l+1}^n\}.
\end{equation}
Thereby,
\begin{equation}
\begin{split}
    &P(\Tilde{\Gamma}_l \geq \Gamma_\text{min} + \alpha)
      = P\left[\Tilde{\Gamma}_l \geq \Gamma_l + \inf \{\Gamma_{l+1}^n\} + \alpha \right] \\
     & = P\left[\Tilde{\Gamma}_l - \Gamma_l - \alpha - \inf \{\Gamma_{l+1}^n\} \geq 0 \right] \\
     &= \sum_{\mu} P(\Tilde{\Gamma}_l - \Gamma_l - \alpha = \mu)P(\inf \{\Gamma_{l+1}^n\} \leq \mu) \\
     & \leq \sum_{\mu} P(\Tilde{\Gamma}_l - \Gamma_l - \alpha = \mu)2^{-r_0 \mu} = \mathbb{E}\left[2^{-r_0( \Tilde{\Gamma}_l - \Gamma_l - \alpha )} \right]\\
     & = 2^{r_0 \alpha} 
     \prod_{i=1}^{l}\mathbb{E}\left[2^{-r_0(\Tilde{\gamma}_i - \gamma_i)} \right] \\
     & \leq  2^{r_0 \alpha} 
     \prod_{i=1}^{l} 2^{r_0 [E_0(\frac{1+r_0}{-r_0},W_N^{(i)}) + b_i]}\\
     &= 2^{r_0 \alpha} 
     2^{r_0\sum_{i=1}^{l}[E_0(\frac{1+r_0}{-r_0},W_N^{(i)})+b_i]},
\end{split}
\end{equation}
where the first inequality is an application of Wald's identity and the second inequality is by Lemma \ref{lemma3}.

\hfill\IEEEQEDhere 

%%%%%%%%%%%%%%%%%%%%%%%%%%%%%%%%%%%%%%%%%%%%%%%%%%%%%%%%%%%%%%%%%%%%%%%%%%%%%%%%%%%%%%%%%%%%%%%%%%%%%%%

\subsection{Proof of Theorem 1}
\begin{equation}
\begin{split}
    & P\left[ \Tilde{\Gamma}_l \geq \Gamma_\text{min} + \alpha \right] \\
    &\leq \sum_{n = 0}^{l-1}Pr\left[ \Tilde{\Gamma}_l \geq \Gamma_{n} + \alpha \right] + 
    P\left[ \Tilde{\Gamma}_l \geq \Gamma_{l} + \inf \{\Gamma_{l+1}^n\} + \alpha \right]\\
    &\leq \sum_{n = 0}^{l-1} 2^{-r \alpha} 2^{-r\sum_{i=1}^{l} \left[ E_0(\frac{1-r}{r},W_N^{(i)}) + b_i \right]} \\
    &~~~~ + 2^{r_0 \alpha} 2^{r_0\sum_{i=1}^{l} 
    \left[ E_0(\frac{1+r_0}{-r_0},W_N^{(i)}) + b_i\right]} \\
    & = (l+1) 2^{-r\alpha} 2^{-r \sum_{i =1}^{l}\left[ E_0(\frac{1-r}{r},W_N^{(i)}) + b_i \right]}
\end{split}
\end{equation}
where the first inequality is by the definition of $\Gamma_\text{min}$ and Boole's inequality. The second inequality is by Lemma \ref{lemma5} and \ref{lemma6}.
The last equality is obtained by assuming $r = -r_0$.

\hfill\IEEEQEDhere 

%%%%%%%%%%%%%%%%%%%%%%%%%%%%%%%%%%%%%%%%%%%%%%%%%%%%%%%%%%%%%%%%%%%%%%%%%%%%%%%%%%%%%%%%%%%%%%%%%%%%%%%

\subsection{Proof of Theorem 2}
\begin{equation}
\begin{split}
    & \mathbb{E}[C_1] \leq \sum_{l=1}^{\infty}\sum_{m}\sum_{\theta=1}^{\infty} \mathbb{E}[C_{l,m,\theta}] \\
    & = \sum_{l=1}^{\infty}\sum_{m}\sum_{\theta=1}^{\infty}
    P(\Tilde{\Gamma}_{l,m} > \Gamma_\text{min} + (\theta - 2)\Delta),\\
    &\leq \sum_{l=1}^{\infty}\sum_{m}\sum_{\theta=1}^{\infty}
    (l+1) 2^{-r\alpha} 2^{-r \sum_{i =1}^{l}\left[ E_0(\frac{1-r}{r},W_N^{(i)}) + b_i \right] } \\
    &= \sum_{l=1}^{\infty}\sum_{m}\sum_{\theta=1}^{\infty}
    (l+1) 2^{-r\alpha}\prod_{i=1}^{l} 2^{-r \left[ E_0(\frac{1-r}{r},W_N^{(i)}) + b_i \right] } \\
    & \leq \sum_{\theta=1}^{\infty}2^{-r\alpha} \sum_{l=1}^{\infty}
    (l+1)2^{lR_l} \prod_{i=1}^{l} 2^{-(R_i + \epsilon) } \\
    &\leq \sum_{\theta=1}^{\infty}2^{-r\alpha} \sum_{l=1}^{\infty}
    (l+1) \left[ 2^{-\epsilon }\right]^{l} 
    \leq \sum_{\theta=1}^{\infty}2^{-r\alpha} \frac{1}{(1-2^{-\epsilon})^2} \\
    & = \sum_{\theta=1}^{\infty}2^{-r(\theta - 2)\Delta} 
    \frac{1}{(1-2^{-\epsilon})^2}
    = \frac{2^{r\Delta}}{1-2^{-r\Delta}}\frac{1}{(1-2^{-\epsilon})^2},
\end{split}
\end{equation}
for $\alpha = (\theta - 2)\Delta$.
The second inequality is by Theorem \ref{theorem 1}. 
The third inequality is by  (\ref{partial rate}) and the upper bound on the number of incorrect nodes at depth $l$ of the decoding tree. 
The infinite sigma on the depth of the tree $l$ is convergent if and only if $\epsilon$ is positive as we assumed it.

The value $\Delta = \frac{1}{r}$ minimizes the upper bound. 
So we have
\begin{equation}
    \mathbb{E}[C_1] \leq  
    \frac{4}{(1-2^{-\epsilon})^2}.
\end{equation}

\hfill\IEEEQEDhere 

%%%%%%%%%%%%%%%%%%%%%%%%%%%%%%%%%%%%%%%%%%%%%%%%%%%%%%%%%%%%%%%%%%%%%%%%%%%%%%%%%%%%%%%%%%%%%%%%%%%%%%%
%%%%%%%%%%%%%%%%%%%%%%%%%%%%%%%%%%%%%%%%%%%%%%%%%%%%%%%%%%%%%%%%%%%%%%%%%%%%%%%%%%%%%%%%%%%%%%%%%%%%%%%

\subsection{Proof of Theorem 3}
For $\beta>1$, probabilties $Q_j$, and a set of nonnegative numbers $a_{jk}$, the Minkowski inequality (MI) is as
\begin{equation}
    \left[\sum_j Q_j \left(\sum_k a_{jk} \right)^{\beta} \right]^{1/\beta}
    \leq \sum_{k} \left(\sum_j Q_j a_{jk}^{\beta} \right)^{1/\beta}.
\end{equation}
By (\ref{expected C0}) and using Minkowski inequality we get the result of
\begin{equation}
\begin{split}
    & (\mathbb{E}[C_1^{\beta}]) ^{1/\beta}
     \leq \left[ \mathbb{E}\left(\left[\sum_{l=1}^{\infty}\sum_{m}\sum_{\theta=1}^{\infty} C_{l,m,\theta} \right]^{\beta}\right) \right]^{1/\beta}\\
    & \overset{\text{MI}}{\le} 
    \sum_{l=1}^{\infty}\sum_{m}\sum_{\theta=1}^{\infty} \left( \mathbb{E} \left[ (C_{l,m,\theta})^{\beta} \right]\right)^{1/\beta}.
\end{split}
\end{equation}

Because $C_{l,m,\theta}$ is a random variable with Bernoulli distribution, $(C_{l,m,\theta})^{\beta} = C_{l,m,\theta}$. Similar to the proof of the previous theorem we have

\begin{equation}
\begin{split}
    & (\mathbb{E}[C_1^{\beta}]) ^{1/\beta} \leq 
    \sum_{l=1}^{\infty}\sum_{m}\sum_{\theta=1}^{\infty} \left( \mathbb{E} \left[ (C_{l,m,\theta})^{\beta} \right]\right)^{1/\beta}\\
    &= \sum_{l=1}^{\infty}\sum_{m}\sum_{\theta=1}^{\infty}P\left[ \Tilde{\Gamma}_l \geq \Gamma_\text{min} + (\theta - 2)\Delta \right]^{1/\beta} \\
    &\leq \sum_{l=1}^{\infty}\sum_{m}\sum_{\theta=1}^{\infty}
    (l+1)^{1/\beta} 2^{-\frac{r\alpha}{\beta} } 2^{-\frac{r}{\beta} \sum_{i =1}^{l}\left[ E_0(\frac{1-r}{r},W_N^{(i)}) + b_i \right] } \\
    &= \sum_{l=1}^{\infty}\sum_{m}\sum_{\theta=1}^{\infty}
    (l+1)^{1/\beta} 2^{-\frac{r\alpha}{\beta}}\prod_{i=1}^{l} 2^{-\frac{r}{\beta} \left[ E_0(\frac{1-r}{r},W_N^{(i)}) + b_i \right] } \\
    & \leq \sum_{\theta=1}^{\infty}2^{-\frac{r\alpha}{\beta}} \sum_{l=1}^{\infty}
    (l+1)^{1/\beta}
    2^{lR_l} \prod_{i=1}^{l} 2^{-( \frac{R_i + \epsilon}{\beta} ) } \\
    &\leq \sum_{\theta=1}^{\infty}2^{-\frac{r\alpha}{\beta}} \sum_{l=1}^{\infty}
    (l+1)^{1/\beta}
    \left[ 2^{\frac{-\epsilon}{\beta} }\right]^{l} 
    \leq \sum_{\theta=1}^{\infty}2^{-\frac{r\alpha}{\beta}} \frac{1}{(1-2^{ \frac{-\epsilon}{\beta} })^2} \\
    & = \sum_{\theta=1}^{\infty}2^{\frac{-r(\theta - 2)\Delta}{\beta}} 
    \frac{1}{(1-2^{\frac{-\epsilon}{\beta}})^2} 
    = \frac{2^{\frac{r\Delta}{\beta}}}{1-2^{\frac{-r\Delta}{\beta}}}
    \frac{1}{(1-2^{\frac{-\epsilon}{\beta}})^2} .
\end{split}
\end{equation}
The value of $\beta$ should be chosen such that the following condition is satisfied.
\begin{equation}
    \beta lR_l \leq \sum_{i =1}^{l}R_i.
\end{equation}
The value of $\Delta = \frac{\beta}{r}$ threshold spacing will minimize the upper bound and we can have
\begin{equation}
     (\mathbb{E}[C_1^{\beta}]) ^{1/\beta} \leq 
     \frac{4}{(1-2^{\frac{-\epsilon}{\beta}})^2}.
\end{equation}
Due to the symmetry of the problem, the same bound is valid for any node.
Finally, using the generalized Chebyshev-inequality, we have
\begin{equation}
    P(C_n \geq L) \leq \frac{\mathbb{E}[C_n^{\beta}]}{L^{\beta}}.
\end{equation}

\hfill\IEEEQEDhere 
%%%%%%%%%%%%%%%%%%%%%%%%%%%%%%%%%%%%%%%%%%%%%%%%%%%%%%%%%%%%%%%%%%%%%%%%%%%%%%%%%%%%%%%%%%%%%%%%%%%%%%%

\bibliographystyle{IEEEtran}


\end{document}